\documentclass{pinchcr}   
\usepackage{amsmath,amssymb,bm}
\newcommand{\be}{\begin{equation}}
\newcommand{\ee}{\end{equation}}

\title{Lectures on nonequilibrium active systems}

\subtitle{Lecture Notes for Les Houches 2018 Summer School on Active Matter and Nonequilibrium Statistical Physics}

\author{L. Berthier}

\affiliation{Laboratoire Charles Coulomb (L2C), Universit\'e de
Montpellier, CNRS, Montpellier, France}

\author{J. Kurchan}

\affiliation{Laboratoire de Physique Statistique, D\'epartement de physique de l'ENS, Ecole Normale Sup\'erieure, PSL Research University, Universit\'e Paris Diderot, Sorbonne Paris Cit\'e, Sorbonne Universit\'es, UPMC Univ. Paris 06, CNRS, 75005 Paris, France}

\begin{document}

\maketitle

\preface

These notes are based on lectures given during the Summer School 
``Active matter and non-equilibrium statistical physics'', held in Les Houches, August 27 - September 21, 2018, organised by G. Gompper, M. C. Marchetti, J. Tailleur, and J. Yeomans. 
During the school, we have shared a series of lectures on ``Active glassy materials'', where we covered topics on active matter, non-equilibrium statistical mechanics, and dense glassy materials. In these notes, we have merged our lectures into a single chapter broadly dedicated to ``Non-equilibrium active systems''. 
We start with a discussion of generic features of non-equilibrium statistical mechanics, followed by a description of selected examples of the possible consequences of not being at thermal equilibrium. We then introduce the topic of dense glassy materials with a short review of glassy dynamics, rheology and jamming transitions for systems that are not active. We then discuss dense active materials, from simple mean-field theories to numerical models and experimental realisations. Finally, we discuss two examples of materials driven out of equilibrium by an oscillatory driving force.  

\medskip

\acknowledgements

We thank the organisers for the invitation to lecture in Les Houches, and the students in the school for the interesting questions and feedback during the lectures.  
We also thank our collaborators E. Flenner, G. Szamel, E. Tjhung for work directly related to the content of these lectures. LB would like to thank more particularly E. Flenner and G. Szamel for their collaboration on a recent review paper~\cite{activereview} that is also partly based on these Lectures.  
The research leading to these results has received funding from the Simons Foundation (\#454933, Ludovic Berthier, \#454943 Jorge Kurchan). 

\tableofcontents

\maintext
\chapter{Equilibrium statistical and mechanics}

\label{sec:intro}

When introducing  Statistical Mechanics, a series of arguments is given to justify the passage from
a detailed microscopic to a statistical approach. Although not explicitly stated, they are of two, rather independent  kinds.

{\bf Macroscopicity:} 
 Here the line of reasoning is the following: in a system composed of molecules, but also of grains of sand and of bacteria,
 the amount of particles is enormous. A detailed description of the fate of each particle is not only impractical but useless, since this is not
 the information we are seeking when we deal with  ensemble properties.
  A macroscopic description is then necessary. It will make use of the geometry of the high dimensionality of phase  space that  leads to
  notions of extensivity and the introduction of non-fluctuating quantities in the thermodynamic limit. This ultimately  leads to the discovery of phase transitions, symmetry breaking, scaling
  and universality.  One may derive in practice all these features starting from the exact dynamics, analytically  in schematic models and approximations,
  and numerically in realistic systems.

{\bf Equilibrium:}
Systems undergoing Hamiltonian dynamics, or dynamics in contact with  suitable equilibrium thermal baths (more about these below)
possess  a property that makes them very special. In Hamiltonian dynamics it is that if one considers a set of initial conditions that cover the energy surface
uniformly, subsequent dynamics preserves uniformity. In contact with a thermal bath, a similar statement can be made with the canonical distribution.
It turns out that these are all manifestations of a {\it time-reversal symmetry}, which makes the dynamics very particular.
In this way, we obtain a number of strong properties:
\begin{itemize}
\item Equilibrium (i.e. microcanonical or canonical measure) allows us to disregard time and compute the values of {\it instantaneous} observations
without having to solve dynamics. If we are interested in correlations between two or more times, as we need for computing the transport coefficients,   then
a dynamical calculation is inescapable: two systems having the same equilibrium may have very different viscosities.

\item Getting rid of time, and thus working with one dimension less, is not the main consequence of equilibrium. More importantly, it gives us
intuition on the roles played by energy and entropy,
inequalities and theorems on irreversibility. These are the principles that forbid a `particle trap' to work, ratchets to turn spontaneously and different thermometers to give differing measures.
\end{itemize}

Active matter and granular matter are macroscopic systems, and we may thus expect them to have all the plethora of emergent phenomena associated with 
large numbers. On the other hand, they do not have any form of time-reversal, so that  the principles associated with equilibrium  do not apply. 
In these first two sections we discuss the time-reversal symmetry, its consequences, and how they fail when it is broken.

\section{An example of a heat bath.}

Let us review briefly the construction of a heat bath that leads us to a generalized Langevin equation (\cite{Zwanzig}, here
we follow \cite{leticia}). 
We start by considering a system coupled to a bath of $M$ harmonic oscillators , whose characteristics we shall
discuss later. For simplicity we assume the system has one degree of freedom, the generalization to more does not bring any new feature.
The Hamiltonian of  system, bath and interaction is:
\begin{equation}
H_{tot}= H_{sys}+H_{bath}+H_{int}+H_{counter}
\label{uno}
\end{equation}
\begin{equation}
H_{sys}= \frac{p^2}{2m} + V(q) \;\;\; ; \;\;\; 
H_{bath}= \sum_1^M \frac{\pi_a^2}{2} + \frac{\omega_a^2}{x_a^2} \;\;\; ; \;\;\; 
H_{int} = q \sum_1^M c_a x_a
\label{cuatro}
\end{equation}
We wish that when we integrate away the bath, we are left with the partition function of the isolated system.
This will happen if we fix the counter-term so that:
\begin{eqnarray}
 & & \int dq_\alpha dp_\alpha  e^{-\beta(H_{bath}+H_{int})} 
=  e^{+\beta H_{counter}} 
\label{cinco}
\end{eqnarray}
which
implies that we have to set $
H_{counter} =  \frac 12 \sum_1^M \frac{c_a^2}{\omega_a^2} \; q^2 $.
The equations of motion of bath and system variables are then:
\begin{equation}
\dot q = \frac{p}m \;\;\; ; \;\;\; \dot p = -V'(q) - \sum_1^M c_a x_a - \sum_1^M   \frac{c_a^2}{\omega_a^2} \; q
\label{motion}
\end{equation}
and
\begin{equation}
\dot x_a = \pi_a \;\;\; ; \;\;\; \dot \pi_a - \omega_a^2 x_a - c_a q(t)
\label{motion1}
\end{equation}
We may immediately solve for the bath variables, in term of its initial conditions and of the system's variables:
\begin{equation}
x_a(t) = x_a(0) \cos(\omega_a t) + \frac{\pi_a(0)}{\omega_a} \sin(\omega_a t) - \frac{c_a}{\omega_a} \int dt' \sin[\omega_a(t-t')]\; q(t')
\label{osci}
\end{equation}
If we plug this into (\ref{motion}), we obtain an effective equation for the system
\begin{equation}
m \ddot q = -V'(q) +\eta(t) - \int_0^t  dt' \Gamma(t-t') \dot q(t') 
\label{lan}
\end{equation}
with  the `memory kernel':
\begin{equation}
\Gamma(t-t') \equiv \sum_1^M \frac{c_a^2}{\omega_a^2 }\cos[\omega_a(t-t')]
\label{gamma}
\end{equation}
and the `noise term':
\begin{equation}
\eta(t)\equiv -\sum_1^M c_a \left[ \frac{\pi_a}{\omega_a} \sin(\omega_at) + \left(q_a(0)+\frac{c_a q(0)}{\omega_a^2}\right)\cos(\omega_a t)\right]
\label{gamma1}
\end{equation}

We now have to specify the characteristics of the bath. If we assume that the oscillators are drawn from a Gibbs distribution with temperature $T$,
the noise becomes random through its dependence on the initial conditions.  Taking into
account the counter-term, it is a simple exercise 
to show that:
\begin{equation}
\langle \eta(t) \rangle =0 \;\;\; \;\;\; ; \;\;\;\;\;\; \langle \eta(t) \eta(t') \rangle = T \; \Gamma(t-t')
\label{fdtsec}
\end{equation}
The noise correlation is directly related to the friction kernel, a property which is sometimes called `fluctuation-dissipation theorem of the second kind'. 
We shall justify later the origin of this name.
Equation (\ref{lan}), supplemented with (\ref{fdtsec}) (and no mention to any oscillators) is known as the `generalized Langevin equation'. The generalization to $N$ degrees of freedom is straightforward.

Two simplifications are often considered. Suppose the function $\Gamma(t)$ is very peaked, and has integral $=2$, so that $\Gamma(t) \rightarrow 2 \delta(t)$.
In this Markovian limit, equations (\ref{lan}),  (\ref{fdtsec}) become, now directly for $N$ degrees of freedom:
\begin{eqnarray}
m \ddot q_i + \gamma \dot q_i + \frac{\partial V} {\partial q_i} + f_i
         &=& \eta_i(t)  \;\;\;\;\; ; \;\;\; \;\; \langle \eta_i(t) \eta_j(t') \rangle = 2T \delta(t-t') \delta_{ij}
\label{lan3}
\end{eqnarray}
Note that only half of the $\Gamma(t)$ enters the friction kernel, because of the time limits.  
In some cases, we may consider the overdamped limit:

\begin{equation}
 \dot q_i =-\left(\frac{\partial V}{\partial q_i}+f_i\right) + \eta_i  \;\;\;\;\; ; \;\;\; \;\; \langle \eta_i(t) \eta_j(t') \rangle = 2T \delta(t-t') \delta_{ij}
 \label{lan2}
\end{equation}
In equations (\ref{lan3},\ref{lan2}) we have added explicitly a force $f_i$ that does not derive from a global potential, either because $\frac{\partial f_i}{\partial f_j} \neq \frac{\partial f_j}{\partial f_i}$ or because the space is not simply connected,
such as a ring or a torus, and the forces `go around'. 
Of course the splitting between `gradient' and `non-gradient' is not unique.

\section{Kramers and Fokker-Planck equations}

For the Markovian versions (\ref{lan3}) and (\ref{lan2}) it is a standard exercise \cite{six} to go from a description
in terms of a stochastic equation to one in terms on the evolution of a probability density.
In the case with inertia (\ref{lan2}) we obtain Kramers' equation: 
\begin{equation}
\dot P({\bf q},{\bf p},t) = -H_{K}P({\bf q},{\bf p},t) \;\;\; ; \;\;\; P({\bf q},{\bf p},t) = e^{-t H_K}P({\bf q},{\bf p},0)
\label{kr}
\end{equation}
with
\begin{equation}
H_K= { \frac{\partial {\mathcal H}}{\partial p_i}
\frac{\partial }{\partial q_i}- \frac{\partial {\mathcal H}}{\partial
q_i} \frac{\partial }{\partial p_i}
} \;\;
\;\; {- \gamma \frac{\partial }{\partial p_i}\left(T
\frac{\partial }{\partial p_i} + \frac{p_i}{m} \right)} \;\;\;
{ - \frac{\partial }{\partial p_i} f_i({\bf q})
}
\label{Hk}
\end{equation}
(summation convention), where we recognise the Poisson bracket
associated with Hamilton's equations, plus a bath, and (possibly) a
forcing term.
The probability density is defined in phase-space.

For the overdamped case, we obtain the Smoluchowski / Fokker-Planck equation for evolution in {\it configuration} space:
\begin{equation}\frac{dP}{dt} = \sum_i
\frac{\partial }{\partial q_i}\left[ T \frac{\partial }{\partial q_i}+
\frac{\partial V}{\partial q_i}+f_i\right]P 
\end{equation}
with
\begin{equation}
\dot P({\bf q},t) = -H_{FP}P({\bf q},t)\;\;\; ; \;\;\; P({\bf q},t) = e^{-t H_{FP}}P({\bf q},0)
\label{fp}
\end{equation}
where we have defined the generator:
\begin{equation}
H_{FP}=-\sum_i \frac{\partial }{\partial q_i}\left[ T \frac{\partial
}{\partial q_i}+ \frac{\partial V}{\partial q_i}+f_i\right]
\label{Hfp}
\end{equation}

\section{Time reversal symmetry, FDT}

The Langevin equation in all its forms possesses, if all forces derive from a potential, a time reversal symmetry.
The simplest version of this is for the Fokker-Planck equation:
\begin{equation}
 e^{\beta V} e^{-tH_{FP} }  e^{-\beta V}= e^{-tH^\dag_{FP} } \;\;\; \forall t \;\;\;
\;\;\; \rightarrow \;\;\;\;\;\;\;\;\;
 e^{\beta V} H_{FP}   e^{-\beta V}= H^\dag_{FP}
\label{prehermi} 
\end{equation}
Using this and the r.h.s. of (\ref{fp}) we obtain
 detailed balance property,  a relation between the probabilities
of going  from a configuration $a$ to a configuration $b$ and
vice-versa:
\begin{equation}
e^{-\beta V(a)} P_{a \rightarrow b} = e^{-\beta V(b)} P_{b \rightarrow a}
\label{detailed}
\end{equation}
The name `detailed' comes from the fact that if we only ask for the
Gibbs-Boltzmann distribution to be stationary, we need only that
(\ref{detailed}) holds added over configurations, and not term by
term.  We can telescope (\ref{detailed}) to obtain for a chain of
configurations:
\begin{equation}
 P_{a_1 \rightarrow a_2}  P_{a_2 \rightarrow a_3} \dots 
 P_{a_{m-1} \rightarrow a_m}= e^{-\beta [V(a_m)-V(a_1)]} 
P_{a_m \rightarrow a_{(m-1)}}  \dots  P_{a_3 \rightarrow a_2}
  P_{a_2 \rightarrow a_1}  
  \label{reverse}
\end{equation}
which means that 
\vspace{.2cm}
\begin{center}
{\bf Probability [path]}= $\;\;
e^{-\beta[V({\mbox{final}})-V({\mbox{initial}})]}\;\;$ {\bf
  Probability [reversed path]}
\end{center}
\vspace{.2cm}
In other words, we have the Onsager-Machlup reversibility:
 the probability of any path going from {\bf $a$} to {\bf $b$} is
  equal to the probability the time-reversed path, times a constant
  that only depends on the endpoints {\bf $a,b$}.

We can also consider a process which has inertia and its energy is, for example,  ${\mathcal H}=\sum_i \frac{{\bf p_i}^2}{2m_i} + V({\bf q})$.  In this
case, something like detailed balance holds, but on the condition that
we reverse also the velocities:
\begin{equation}
e^{-\beta {\mathcal H}(a)} P_{a \rightarrow b} = e^{-\beta {\mathcal
    H}(\bar b)} P_{\bar b \rightarrow \bar a}
\label{detailedk}
\end{equation}
where $\bar a$, $\bar b$ are the configurations $a,b$ with the
velocities reversed.

\section{Equilibrium theorems: stationary distribution, reciprocity and Fluctuation-Dissipation theorem (FDT)}

\subsection{Stationarity and equilibrium \label{sec}} 

A first property of detailed balance, or its modified form in the case of inertia, is that the Gibbs-Boltzmann distribution is stationary.
\begin{equation}
\frac{d}{dt} P_{stationary}(q) =0 \;\;\;\; {\mbox{for}} \;\;\;  P_{stationary}(q) = \frac{e^{-\beta V(q)}}Z
\end{equation} the r.h.s holds if  $f_i=0$, i.e. the forces derive from a global potential.
If there is a single stationary state, then:
\begin{equation}
\lim_{t \rightarrow \infty} e^{-tH_{FP}} P(q,t=0) = P_{stationary}(q)
\end{equation}
Similarly, for the Kramers operator if $f_i=0$:
\begin{equation}
\lim_{t \rightarrow \infty} e^{-tH_{K}} P(q,p,t=0) = P_{stationary}(q,p)
\end{equation}
If 
\begin{equation}
 P_{stationary}(q,p) = \frac{     e^{-\beta {\mathcal H} (q,p)}    }Z
\end{equation}
{\it If $f_i \neq 0$, i.e. the forces do not derive from a global potential, then there is no simple {\it a priori}
expression for the distribution. Indeed, in practice beyond one degree of freedom  one has to resort to approximate
schemes.}
 
\subsection{Reciprocity}

Consider  the correlations:
\begin{equation}
\langle B(t)A(t') \rangle = \int dq \left[ B e^{-(t-t')H_{FP}} A e^{-t'H_{FP}}\right] P(q) =C_{BA}
\end{equation}
A time-reversal symmetry leads naturally to a symmetry of times of measurement. 
Using   detailed balance, we have that, if the forces derive from a potential, then:
\begin{equation}
\langle B(t)A(t') \rangle \rightarrow C_{BA}(t-t') = C_{AB}(t-t')
\label{onsager}
\end{equation}
if both $A$ and $B$ depend on coordinates -- a sign coming from time-reversal appears if there are velocities involved.
As  $t' \rightarrow \infty $ we have $C_{BA}(t,t') \rightarrow C_{BA}(t-t')$.
This reciprocity relation becomes more useful when combined with Linear Response results below.
 
\subsection{Response and Fluctuation-Dissipation theorem}

Consider a system perturbed  by a `kick' in $A$:
\begin{equation}
V \rightarrow V + h \xi(t-t')  A
\end{equation}
where $\xi$ has the form of an impulse in $t\sim t'$. 
Compute   the response of $\langle B(t)\rangle$ for small perturbation:
\begin{equation}
R_{BA}(t, t') = \left. \frac{\delta}{\delta h} \langle B(t) \rangle_h \right|_{h=0}
\end{equation}
Still within linear response, the effect of a general time-dependent small field $h(t)$ may be expressed as a superposition of impulses 
\begin{equation}
\langle B(t) \rangle_h = \int dt' \; R_{BA}(t,t') h(t') 
\end{equation}
In particular, the effect of a constant field that is on from $-\infty < t'' <t'$ is:   
\begin{equation}
\langle B(t) \rangle_h = h \int_{-\infty}^{t'} dt'' \; R_{BA}(t,t'') = h \int dq \;  B(q) \; P(q,t;q't') \; \frac{e^{-\beta(V+hA)(q')}}{Z_h}
\label{rhs}
\end{equation}
where we have used linear response on one hand, and the fact that the system is in equilibrium with the field $h$ at time $t'$, on the other.
Developing to first-order for a small perturbation:
\begin{equation}
\frac{e^{-\beta(V+hA})}{Z_h} \sim e^{-\beta V} \left[ 1-\beta h (A-\langle A )\rangle \right]
\end{equation}
and replacing this in the r.h.s. of (\ref{rhs}) we obtain:
\begin{equation}
\langle B(t) \rangle_h - \langle B(t) \rangle = h \int_{-\infty}^{t'} dt' \; R_{BA}(t-t') =  - \beta h \left[ C_{BA}(t-t')-\langle B \rangle  \langle A \rangle \right] 
\end{equation}
where we have also used stationarity in equilibrium $R_{BA}(t,t') \rightarrow R_{BA}(t-t')$.
Differentiating wrt $t'$, we obtain the Fluctuation-Dissipation relation (of the first kind):
\begin{equation}
R_{BA}(t-t') = \beta \frac{\partial C_{BA}}{\partial t'} (t-t') = - \beta \frac{\partial C_{BA}}{\partial t} (t-t')
\label{fdt1}
\end{equation}

\section{ Fluctuation, Dissipation and measurement}

A thermometer is a subsystem that exchanges heat with some physical quantity of the system to be measured. It has to be either small or
weakly coupled, because we wish that the system is not dramatically altered by the measure  process. 
The reading indicated by the thermometer is just a measure of its own energy. Clearly, the fact that a thermometer indicates some temperature
does not require that what is measured  be in thermal equilibrium, as we know from our own medical experience.
 We usually take for granted that any thermometer
coupled to a homogeneous system will indicate the same temperature. This is clearly so if the system is in thermal equilibrium,
but need not be the case when it is not, even if it is macroscopically homogeneous, as we shall see.

\begin{figure}
\begin{center}
\includegraphics[width=5cm]{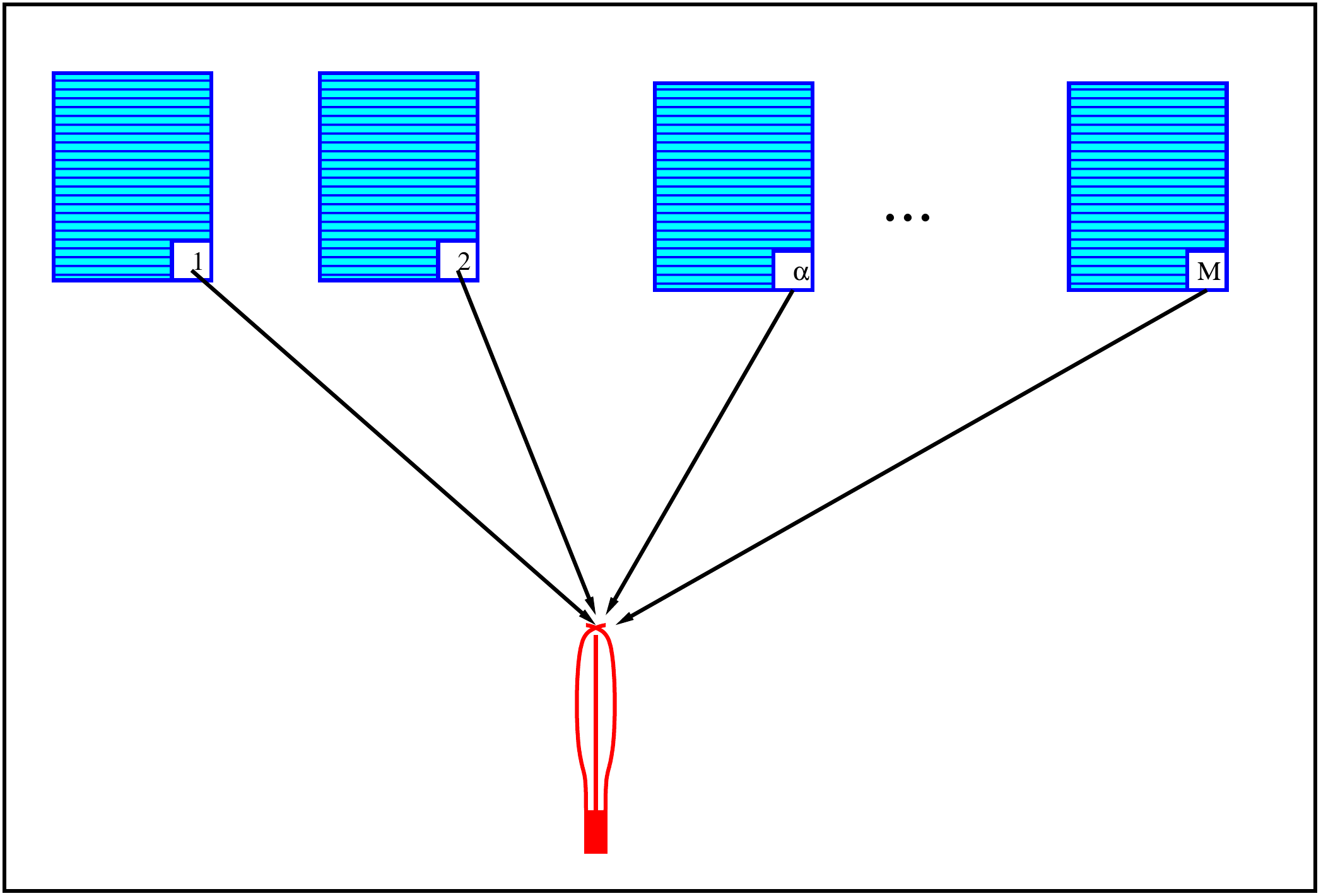}
\vspace{.2cm}
\caption{A sketch of the thermometer described in Eq.~(\ref{eq:sketch}).}
\label{lstar}
\end{center}
\end{figure}

It is interesting to work out  in detail how these questions materialize in a model system~\cite{six,teff}. In order to make the perturbation due to the thermometer
coupled to an observable $A$ the weakest possible, and also in order to make the statistics better, so that the time of measurement may be short,
we shall assume that the thermometer  simultaneously couples to $\alpha=1,...,M$ copies of different realizations of the same system
with coordinates ${\bf q}^\alpha,{\bf p}^\alpha$, see Fig.~\ref{lstar}.
We  intend to use them in turn  as baths for a 
probe of coordinates    ${\bf q}^{th},{\bf p}^{th}$ and energy ${\mathcal
H}^{th}({\bf q}^{th},{\bf p}^{th})$. To do this we couple them, for example through a term:
\begin{equation}
{\mathcal H} = {\sum_\alpha {\mathcal H}}   ({\bf q}^{\alpha},{\bf p}^{\alpha})        +  {
{\mathcal H}^{th}({\bf q}^{th},{\bf p}^{th})} - 
{M^{-\frac{1}{2}} \sum_\alpha {\bf q}^\alpha 
{\bf .} {\bf q}^{th}}
\label{eq:sketch}
\end{equation}
We may ask what is the condition for the coupling term to constitute a 
legitimate thermal bath for the primed system. 
The equations of motion of the thermometer variables is: 
\begin{equation}
\ddot q^{th} = -\frac{\partial \mathcal H^{th}}{\partial q_i^{th}} - {\bf h(t)}
\end{equation}
Where the field ${\bf h}$ is 
$ {\bf h}=M^{-\frac{1}{2}}\sum_\alpha q_\alpha$.
The large $M$ limit allows us to treat each $M^{-\frac{1}{2}}  {\bf q}^\alpha 
{\bf .} {\bf q}^{th} $ as a small perturbation to the system ${\mathcal H}_\alpha$, and  $M^{-\frac{1}{2}} \sum_\alpha {\bf q}^\alpha$ as a Gaussian.
 Assuming without loss of generality that   
$\langle q_\alpha \rangle=0$ in the absence of coupling, the field ${\bf h}$ has two contributions
for large $M$: {\em i)} a random Gaussian noise $\eta(t)$ with correlation $C_{\alpha \alpha}(t,t')=
\langle \eta(t) \eta(t') \rangle = \langle q_\alpha(t) q_\alpha(t')\rangle$ and {\em ii)} a drift
due to the back effect of the $q'$ which acts as a field  on the $q_\alpha$. Again, because
 $M$  is large, the average response of the ensemble $\alpha$ is: 
\begin{equation}
M^{-\frac{1}{2}} \langle q^\alpha \rangle = \int_0^t \;  dt' \; R_{\alpha \alpha}(t,t') q^{th}(t')
\end{equation}
The equation of motion of the thermometer variable becomes:  
\begin{equation}
\ddot q^{th} = -\frac{\partial \mathcal H^{th}}{\partial q^{th}} +\eta(t) + 
\int_0^t \;  dt' \; R_{\alpha \alpha}(t,t') q^{th}(t')
\end{equation}
The thermometer is in fact evolving as in contact with a bath constituded by the $M$ copies of our system,
through  the generalized Langevin equation (\ref{lan}).

Consider now the particular case in which  the systems are in equilibrium. One has the Fluctuation-Dissipation relation:
\begin{eqnarray} 
T R_{\alpha \alpha}((t-t') &=& \frac{\partial}{\partial t'} C_{\alpha \alpha}(t-t') \nonumber\\
T \; {\mbox{Im}} \;  R_{\alpha \alpha}((\omega) &=&  \omega \; C_{\alpha \alpha}(\omega)
\label{fdtalpha}
\end{eqnarray}
the latter in Fourier space.
The thermometer will, under the action of this dynamics, equilibrate to
the Gibbs-Boltzmann distribution\footnote{With an energy that includes a contribution $\langle [q^{\alpha}]^2  
\rangle q^{th \; 2}$ coming
from the interaction term, which we may neglect if the interaction is weak. }.
 Fluctuation-dissipation of the second kind for the $q_\alpha$ has become a fluctuation dissipation
relation of the first kind when they are considered as a bath for the thermometer $q^{th}$.
{\it It follows that the nature of the thermometer is, in this case, immaterial: it will always thermalize to the right temperature}.
Imagine instead that the copies of the system are not in equilibrium, for example because they are
composed of active matter (we shall come to this in following sections). In that case, we will have instead a relation
\begin{eqnarray} 
T(\omega) \; {\mbox{Im}} \; R_{\alpha \alpha}(\omega) = i \omega \;  C_{\alpha \alpha}(\omega)
\label{fdtalpha1}
\end{eqnarray}
which, in fact, defines a non-constant $T(\omega)$. Now, it is easy to estimate what will happen with 
our thermometer $q^{th}$: if its structure is such that it responds to higher frequencies, it will `see' a bath of 
$T(\omega)$ for large $\omega$ and,  if it responds to lower frequencies, for $T(\omega)$ for lower $\omega$.
It is hence clear that different thermometers will respond differently, depending on their own internal dynamics.

\chapter{Out of equilibrium and `Why Not?' questions}

\label{sec:jorgewhynot}

\section{Five ways to get out of equilibrium}

Let us summarize some ways to set a system out of equilibrium.  In all these cases the Gibbs-Boltzmann 
distribution is generically lost, and there is no general, explicit expression that will give the distribution in terms of the forces without
actually solving for the dynamics.

\vspace{.2cm}

$\bullet$ {\bf Many baths. Non equilibrium steady states. A model for active matter}

\vspace{.2cm}

The most obvious way  is to establish transport between two regions of the sample.
For example, if our system has many variables $X,Y,...$ and $V=V(X,Y,...)$ we may connect  heat baths to $X$ and $Y$ with 
different characteristics:
\begin{eqnarray}
\ddot X &=& - \frac{\partial V}{\partial X} + \eta_1(t) -\int dt' \Gamma_1(t-t') \dot X \;\;\; ; \;\;\; \langle \eta_X(t)\eta_X(t')\rangle = T_1 \Gamma_1(t-t')\nonumber \\
\ddot Y &=& - \frac{\partial V}{\partial Y} + \eta_2(t) -\int dt' \Gamma_2(t-t') \dot Y \;\;\; ; \;\;\; \langle \eta_Y(t)\eta_Y(t')\rangle = T_2 \Gamma_2(t-t')
\label{transport}\end{eqnarray}
If $T_1 \neq T_2$ the system will transport heat from the highest to the lowest temperature, and will thus be out of equilibrium.
A less intuitive form of transport happens when some physical quantity is coupled to a bath that is itself out of equilibrium, i.e. it
does not satisfy Fluctuation-Dissipation itself. The easiest way to see this is to consider two baths with different temperatures as above,
but coupled to the same physical quantity:
 \begin{eqnarray}
\ddot X &=& - \frac{\partial V}{\partial X} + \eta_1(t) -\int dt' \Gamma_1(t-t') \dot X+\eta_2(t) -\int dt' \Gamma_2(t-t') \dot X \nonumber \\
& &\langle \eta_X(t)\eta_X(t')\rangle = T_1 \Gamma_1(t-t')    \;\;\; ; \;\;\;  \langle \eta_Y(t)\eta_Y(t')\rangle = T_2 \Gamma_2(t-t')
\label{transport2}\end{eqnarray}
If $\Gamma_1(t)$ and $\Gamma_2(t)$  have different time-dependences and $T_1 \neq T_2$ , the system will be out of equilibrium, there is transport even at the
level of a single degree of freedom. 
If, for example, we make $T_1>T_2$ and $\Gamma_1$ narrower in time that $\Gamma_2$, we will be injecting energy in the 
high frequencies and dissipating in the lower ones. This is very much like a model for active matter, where the system is coupled to a thermal bath and receives energy at a lower frequency to `propel' itself. We shall use this analogy in Sec.~\ref{twobaths}.

\vspace{.2cm}

$\bullet$ {\bf Forces that do not derive from a global potential}

\vspace{.2cm}

As we mentioned in previous sections, mechanical work will drive a system out of equilibrium if forces do not derive from a potential. This 
is not only the case for forces $f_i$ such that $\frac{\partial f_i}{\partial q_j}-\frac{\partial f_j}{\partial q_i}\neq 0$, but also when  the nontrivial topology of space allows for a local gradient to do work, as is the case of a constant field around a ring.

\vspace{.2cm}

$\bullet$ {\bf Time-dependent drive, Floquet systems.}

\vspace{.2cm}

Time-dependent drive will also pump energy into a system, the sense of the flow on average 
is into the system, as required by the Second Principle. Two cases are interesting: a force randomly dependent on time is, from our
point of view, a heat bath with infinite temperature -- it has zero friction kernel. As such, it will give energy to a system whatever
its temperature.  
Another case, much studied in quantum systems, is when forces are periodic in time. If the system is in contact with a thermal bath (without bath it will heat up),
it may reach a periodic `Floquet'  regime. One may observe the system stroboscopically: then the system looks invariant,
but will definitely be out of equilibrium.

\vspace{.2cm}

$\bullet$ {\bf Initial conditions, slow dynamics.}

\vspace{.2cm}

Glasses, systems with defects and  quasi-integrable systems have very long relaxation times, even when connected
to a `good' equilibrium thermal bath. They are  thus out of equilibrium throughout the observation time.
The question is in turn more complex in dense active matter systems, as we will discuss in section V, which are both driven (and cannot reach any kind of equilibrium) and `glassy' (reaching a steady state is very slow).
 
\vspace{.2cm}
 
$\bullet$ {\bf Conditioned measures, large deviations.}

\vspace{.2cm}

A somewhat more abstract form of non-equilibrium does not depend on the system but on the manner of observation.
In some cases we are interested in observing  equilibrium systems on time intervals and sub-ensembles chosen so that something 
unusual happens. Once we condition the measurement to what the system is doing, we are biasing the measure and
are probing an out of equilibrium situation, even if the system is not perturbed.
As an example consider an equilibrium, supercooled liquid. We know that such a system has regions that decorrelate much faster 
than the rest. If we concentrate on those regions, and for example attempt to check fluctuation-dissipation, 
there is no reason why the equilibrium  relation should hold. 

\section{Equilibrium and out of equilibrium distributions}

\begin{figure}
\begin{center}
\includegraphics[width=5cm]{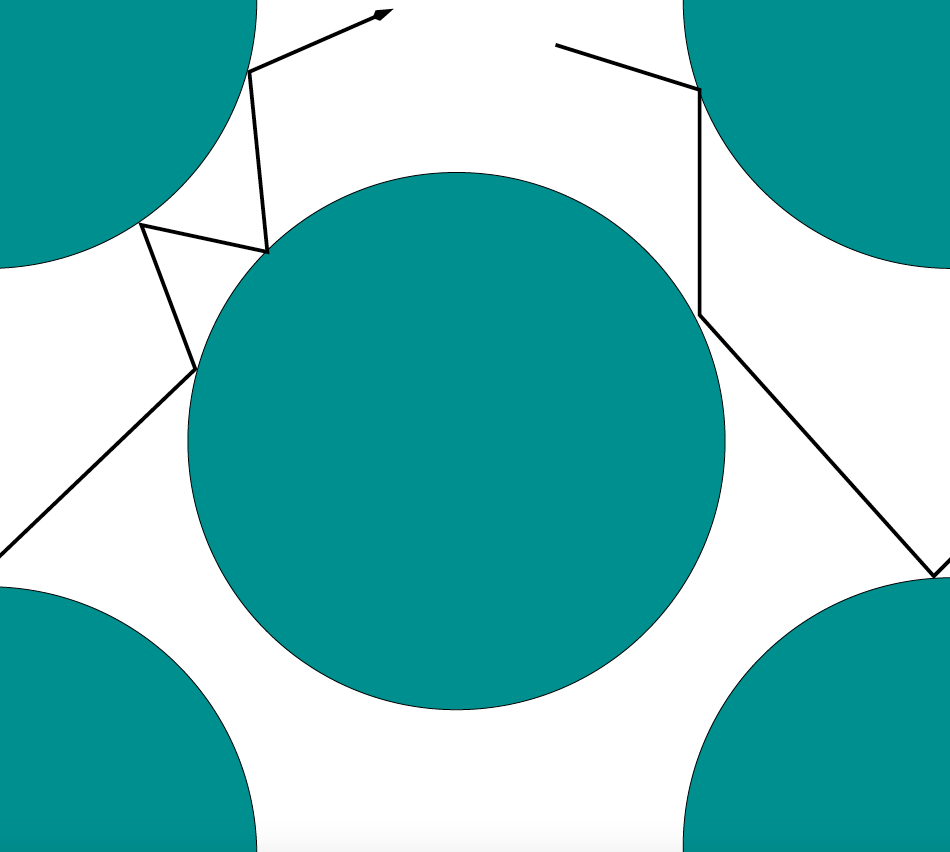}\hspace{1.5cm}
\includegraphics[width=5cm]{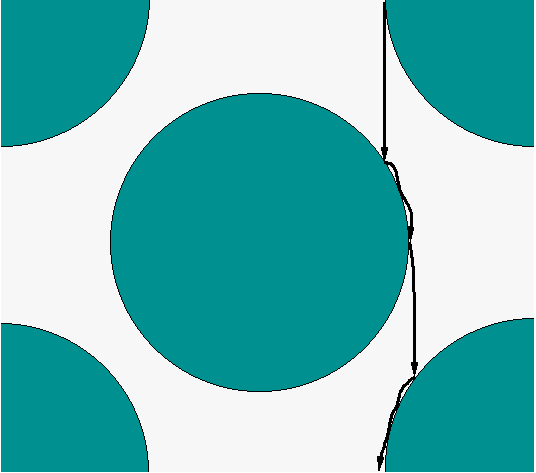}
\vspace{.2cm}
\caption{
Trajectories of the Sinai billiard with periodic boundary conditions, in and out of equilibrium. In the figure
on the right, there is a constant field pointing downwards.}
\label{lstar2}
\end{center}
\end{figure}

As mentioned in Sec.~\ref{sec:intro}, equilibrium statistical mechanics relies on a very specific property of Hamilton's equations
(or its quantum counterpart) that guarantees that a `flat' distribution over the energy surface is preserved.
Once we drive the system, for example with nonconservative forces, we lose energy conservation. This in itself may be 
compensated by a thermal bath. We could even write a variant of the Langevin equation (\ref{lan1}) that will guarantee strict conservation of energy (a Hoover thermostat \cite{hoover}):
 \begin{equation}
m \ddot q_i + \gamma(t)  \dot q_i + \frac{\partial V} {\partial q_i} + f_i=0
\label{lan1}
\end{equation}
where there is no noise and the friction $\gamma(t)$ is time-dependent and tuned so as to keep energy constant, this
will be so if:   $\gamma(t)= -\frac{\sum f_i \dot q_i}{\dot q_j^2}$.
So now we have a driven system that conserves energy, can it be possible that a microcanonical measure correctly describes it?
In fact the answer is negative: as soon as we force the system the measure concentrates on a lower dimensional subset. 
A simple example is the Lorentz gas (or Sinai's billiard) with periodic boundary conditions, see Fig. \ref{lstar2}. If we do not force the system we obtain an ergodic measure, in fact this is one of the few models for which we know a rigorous proof of ergodicity. If instead we force it with an electric field, 
the measure is severely restricted \cite{six}. For very large fields, a trajectory looks like the  figure to the right: almost all of the configuration space is unvisited.
The same applies to the Sinai billiard in contact with a heat bath: even noise will not compensate for the concentration in a subregion of phase space.

Can we derive a distribution that will, at least approximately, allow us to compute instantaneous expectation of observables
in a nonequilibrium steady state,  or periodic situation, given that a `flat' (or Gibbs-Boltzmann) distribution does not faithfully represent a driven system? One strategy that immediately comes to mind~\cite{Jaynes} is to consider a distribution that is flat -- or maximizes entropy --
conditioned to some constraints that we know {\it a priori}. For example, in the case of a system that has a given average energy
and is known to be transporting current at a given rate, we could envision considering a flat average over the subspace of the energy surface  such that the average current is given.
If we did this with the driven Sinai billiard above, we would be restricting to a given value of velocity component $p_y$ and
to a kinetic energy $p_x^2+p_y^2$, but all allowed values of $(x,y)$ within the billiard would have the same weight. Looking at Fig. \ref{lstar2} we immediately see that this is wrong, although less wrong than the uniform microcanonical distribution.  
In fact, this `MaxEnt' approach  gives bad results  and misses important physics  for generic out of equilibrium thermodynamic systems (see discussion in \cite{aurell}). However, in some cases a MaxEnt approach may be a good approximation, but when this happens it requires an explanation.
An approach of this kind was proposed by Edwards \cite{edwards} for granular matter, that has been shown to be a reasonable approximation in certain cases.

\section{`Why Not?' questions}

\vspace{.2cm}

\begin {figure}
\begin{center}
\includegraphics[height=3cm]{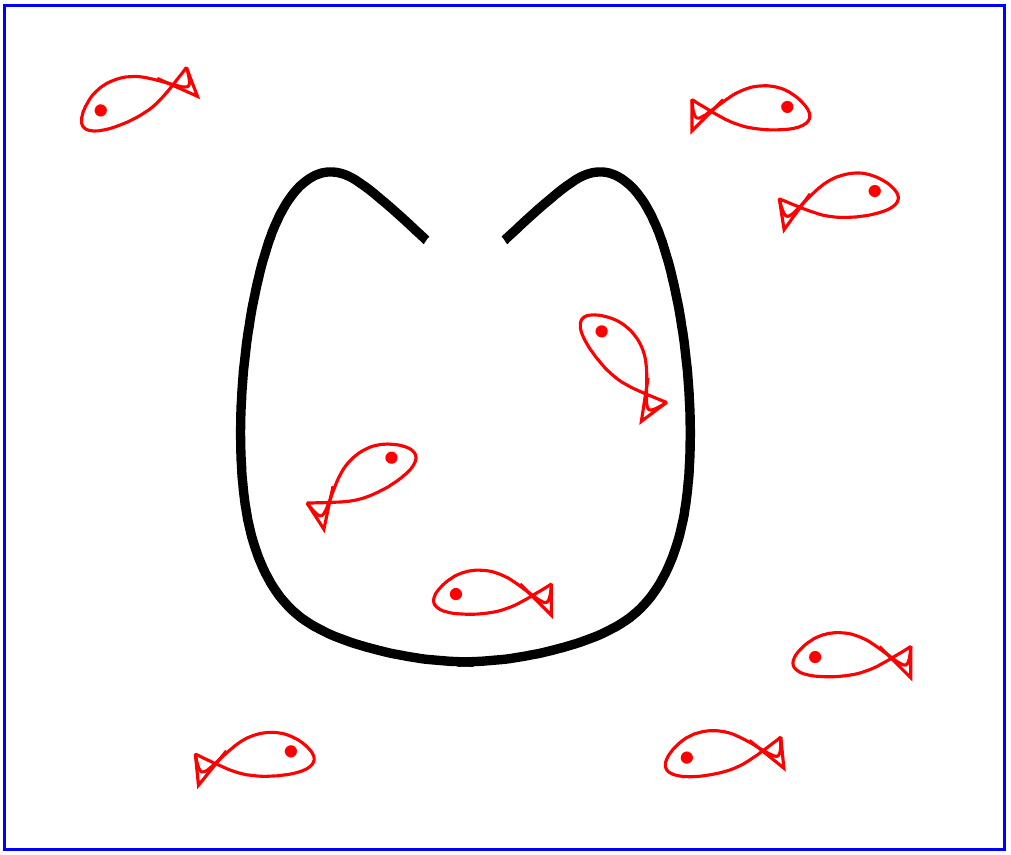} \hspace{.5cm}
\includegraphics[height=3cm]{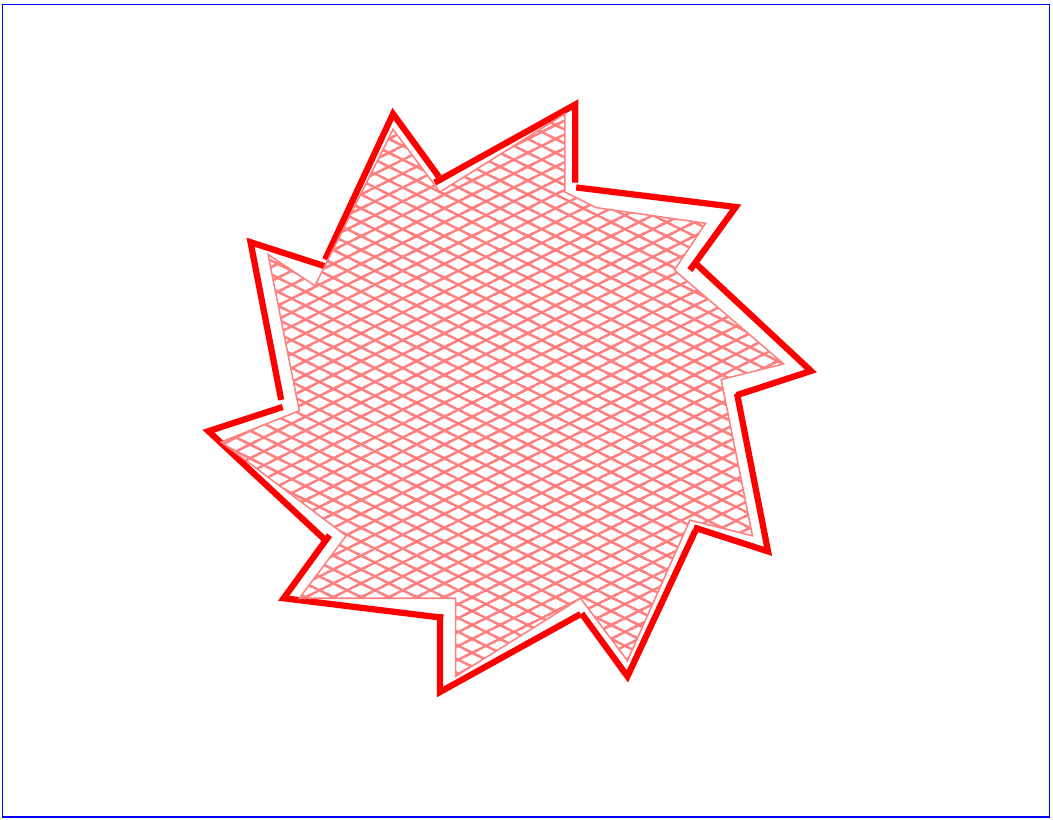}\\
\vspace{1cm}
\includegraphics[height=4cm]{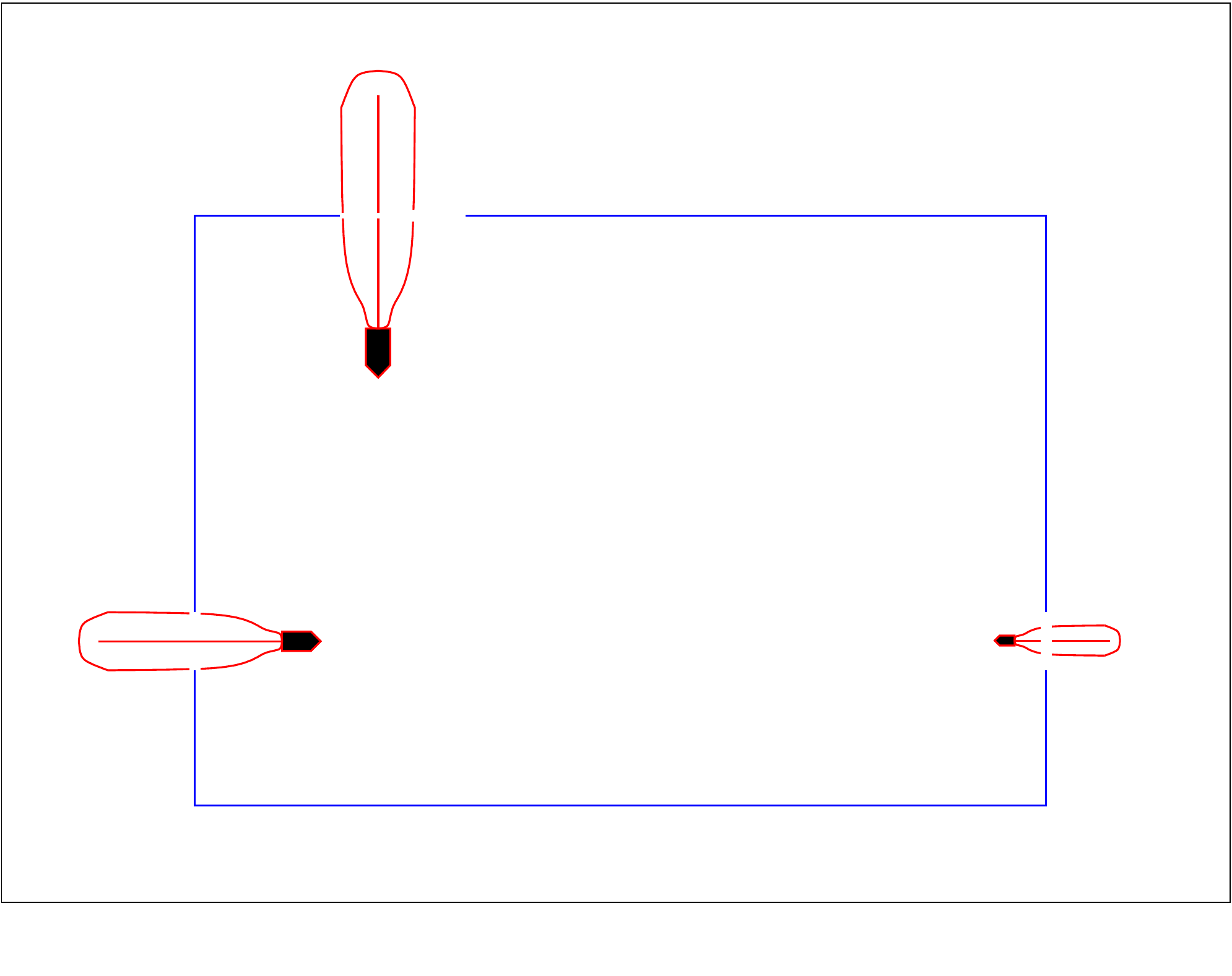}
\caption{"Why not?" questions: (a) is the density of  (noninteracting) fish different inside and outside the trap? (b) Does the ratchet turn? (c) Do different thermometers
indicate  different temperatures?}
\label{whynot}
\end{center} 
\end {figure}

Active matter has the virtue of highlighting, by contrast, the strong constraints posed by equilibrium.
Consider the three examples in Fig.~\ref{whynot}. If the fish were in thermal equilibrium and non-interactive (e.g. small and dead), detailed balance applies.
Choose a point inside and a point outside the trap and apply detailed balance: the probability of getting trapped is equal to the probability of escaping.
The same can be said of the entire trajectory leading in or out.
Similarly, the ratchet in contact with a liquid in equilibrium has equal probabilities of following a given angle vs. time trajectory (see formula (\ref{reverse}), and its reverse. Hence, on average,
there is no net rotation.  Finally, as we saw in Sect. 1.5, we expect three different thermometers to indicate the same temperature, mainly because 
fluctuations and dissipation of the water have the right relation.

Now we may ask the questions for an out of equilibrium situation: living fish, liquids composed of active particles, shaken granular matter.
  Will the density of fish inside the trap be different from outside, will the ratchet turn in some preferential sense, will the thermometers indicate different temperatures, even if the active liquid is homogeneous?
  The correct answer to these questions is `why not'? A process that is not forbidden by any symmetry or conservation law has zero probability
  of not happening.
  A different sort of question is: is the density of fish higher inside or outside the trap, in what sense does the wheel turn, which thermometer indicates a higher temperature?  These questions cannot be answered in any general, model-independent form. The actual answer depends on the nature of activation, and
  the sign may reverse for the same model upon changing parameters.  
  
  A similar  question to the ones of  thermometers has been asked with manometers   \cite{julien}, and the answer is the same: different manometers measure different pressures. This striking result becomes more acceptable if one thinks that, just as temperature being the conjugate thermodynamic quantity
  to energy, pressure is conjugate to volume.   
  
\chapter{Glassy dynamics and jamming transition}

The topic of this section is a widely observed phenomenon. Take a dense system of `particles', which can be molecules, droplets, cells, grains, or animals. When the density is not too large, these particles can easily move, and they can be fueled by thermal fluctuations, chemical reactions, internal motors, or muscles. The system is in a fluid-like state. As the density increases, it becomes increasingly difficult for the particles to find pathways that allow them to move over large distances. The competition between particle crowding in a dense environment and the energy injected at the particle scale may result in a transition from a fluid regime to a dynamically arrested regime where individual particles are permanently trapped by their neighbors. In this arrested state, the particles respond as a homogeneous block to external perturbations; the system has become a solid. Very simple systems, such as assemblies of identical particles in thermal equilibrium would easily crystallize at large densities, but for many `complex' particles the arrested state is fully disordered. The phase transformation between an equilibrium fluid and an arrested amorphous state is the glass transition~\cite{today}. 

This transition from a fluid to an amorphous solid is ubiquitously observed not only for molecules and small colloids (which form molecular and colloidal glasses), but is similarly relevant to describe a large class of active materials~\cite{review1,review2,review3}, where the `particles' can be phoretic colloids, self-propelled grains, or crawling cells. In those examples, the competition arises between the crowding of the active particles (that tends to arrest them) and the intensity of the active forces (that make them move). 

In the active matter community, the fluid-solid transition in active materials has often been termed a jamming transition (see, for a single example, a recent paper entitled `{\it A fluid-to-solid jamming transition underlies vertebrate body axis elongation}', which in fact deals with glassy dynamics~\cite{Mongera2018}), rather than a glass transition. We see two reasons for this. First, the word jamming itself is perhaps more easily grasped by non-glassy experts. Second, it echoes work performed in the granular matter community about 20 years ago that attempted to unify the physics of seemingly disparate physical systems, from molecules to grains and foams~\cite{jamming2}. In a sense, cells, robots and phoretic colloids would only be additional examples of the same type of physics. 

Recently, however, the distinction between the glass and the jamming transition, and the specific features associated with both phenomena have been clarified and explained in great detail~\cite{francesco}. Broadly speaking, the competition between crowding and particle agitation leads to the glass transition phenomenon. In contrast, jamming is understood as a purely geometric transition between viscous and rigid behavior in the absence of any kind of dynamics. Thus, jamming is a zero-temperature, or, for the purpose of the present paper, a zero-temperature and zero-activity limit. Strictly speaking, therefore, particles with non-vanishing activity cannot undergo jamming.

\section{Short review of equilibrium glass transition} 

\label{sec:equilibriumgt}

Let us first quickly review the main features of the equilibrium dynamics of non-active (thermal, passive) fluids approaching their glass transition. For brevity, we will use the words ``equilibrium glass transition'' to refer to this case. 

The most noticeable phenomenon accompanying the incipient glass transition is the enormous slow down of the dynamics~\cite{BerthierBiroliRMP}. For instance, the viscosity of a hard sphere colloidal glass former can 
increase by seven orders of magnitude when the colloidal suspension's volume fraction changes from a dilute value of a few percent to a value close to the so-called colloidal glass transition~\cite{Cheng2002,Russel2013}. 
Even more impressively, a viscosity of a good molecular glass former can increase by twelve orders of magnitude upon decreasing the temperature by a mere factor of two~\cite{Angell1995}. 

To set the stage for the discussion of active glassy dynamics in Sec.~\ref{active}, we discuss the salient features of 
the structure and dynamics of equilibrium (non-active, \textit{i.e.} `passive') glassy systems. We  
note that many of the microscopic phenomena discussed in this section require detailed information about 
particles' motion on the microscopic scale, and for that reason they were first observed in computer
simulations and later studied in colloidal systems. We also recall that almost all glass-formers
studied in computer simulations are many-component mixtures (for single component systems with typical
interaction potentials it is virtually impossible to avoid crystallization upon even mild supercooling). 

All the examples shown in this section and in the next one were obtained for a 50:50 binary mixture of  spherically symmetric particles interacting via the Lennard-Jones potential cut at the minimum, which is usually
referred to as the Weeks-Chandler-Andersen (WCA) interaction~\cite{WCA},
\begin{equation}
\label{potential}
V_{\alpha \beta}(r) = 4 \epsilon
\left[ \left( \frac{\sigma_{\alpha \beta}}{r} \right)^{12}
- \left( \frac{\sigma_{\alpha \beta}}{r} \right)^6 \right],
\end{equation}
for $r \le \varsigma_{\alpha \beta} = 2^{1/6} \sigma_{\alpha \beta}$
and constant otherwise. In Eq.~(\ref{potential}), $\alpha, \beta$ denote the particle species
$A$ or $B$, $\epsilon =1$ (which sets the unit of energy),
$\sigma_{AA} = 1.4$, $\sigma_{AB} = 1.2$, and
$\sigma_{BB} = 1.0$ (which sets the unit of length). All the figures are adapted from Refs.~\cite{glassline,activereview}.

\begin{figure}
\begin{center}
\includegraphics[scale=0.3]{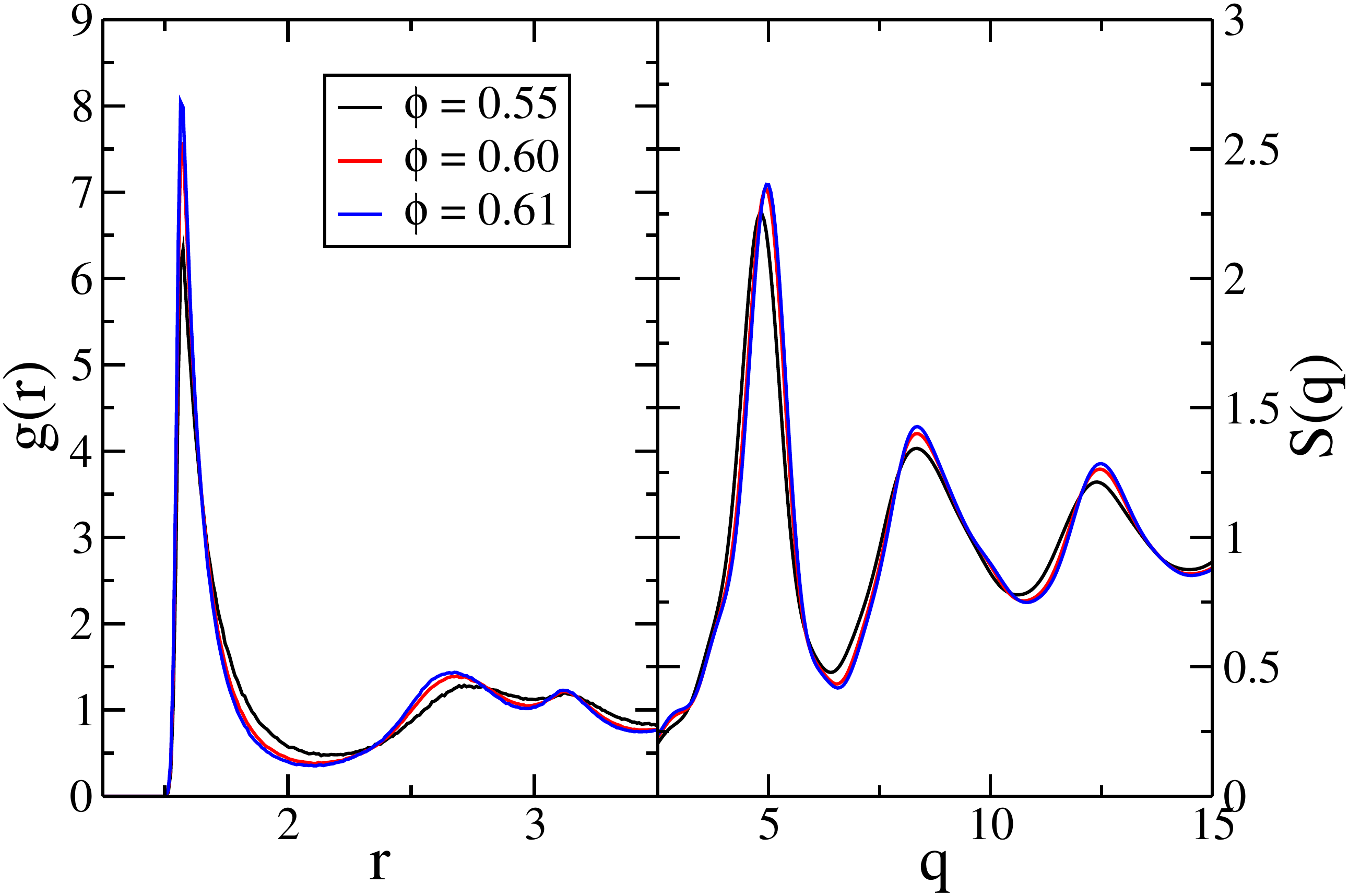}
\caption{\label{fig:pstatics} The pair correlation function, $g(r)$, (left panel) and the static
structure factor, $S(q)$, (right panel), for three volume fractions in the vicinity of the glass transition for 
an equilibrium WCA system at a low temperature, $T=0.01$.}
\end{center}
\end{figure}

A defining feature of an approaching glass transition is a dramatic slowing down of a liquid's dynamics with 
little change of the pair structure upon a small change of the temperature $T$ and/or density $\rho$. The pair 
structure is typically monitored through the pair correlation function~\cite{HansenMcDonald}
\begin{equation}
\label{eq:pair}
g(r) = \frac{1}{\rho N} \left< \sum_{n,m\ne n} \delta[\mathbf{r} - ( \mathbf{r}_n(0) - \mathbf{r}_m(0)) ] \right>
\end{equation}
or the static structure factor~\cite{HansenMcDonald}
\begin{equation}
\label{eq:structure}
S(q) = \frac{1}{N} \left< \sum_{n,m} e^{i\mathbf{q} \cdot (\mathbf{r}_n(0) - \mathbf{r}_m(0))} \right>.
\end{equation}
In Eqs.~(\ref{eq:pair}) and (\ref{eq:structure}),
$N$ is the number of particles and $\mathbf{r}_n(t)$ is the 
position of particle $n$ at a time $t$. While these two functions are related through a Fourier transform [$S(q) = 1 + \rho \int d {\bf r} e^{i {\bf q} \cdot {\bf r}} (g({\bf r})-1)$]
and thus encode the same information,  
it is easier to distinguish differences in structure on nearest neighbor length scales by examining 
$g(r)$ and it is easier to compare the decay of the structure on longer length scales by examining $S(q)$.

In Fig.~\ref{fig:pstatics} we show the density dependence of the pair correlation function and the static structure
factor of a WCA system at the constant temperature, $T=0.01$. While the
pair structure changes very little, the long-time dynamics of the system (as characterized by time-dependent correlation functions defined below) 
slows down by approximately 3 orders of magnitude. 

To determine if the structure evolves in time we can examine time dependent versions of Eqs.~(\ref{eq:pair}, \ref{eq:structure}) where $\mathbf{r}_n(0)$ is replaced by $\mathbf{r}_n(t)$. 
The time dependent version of Eq.~(\ref{eq:structure}) defines the collective (coherent) intermediate scattering 
function~\cite{HansenMcDonald}
\begin{equation}
\label{eq:collective}
F(q;t) = \frac{1}{N} \left< \sum_{n,m} e^{i\mathbf{q} \cdot [\mathbf{r}_n(t) - \mathbf{r}_m(0)]} \right>,
\end{equation}
which characterizes the relaxation of the initial structure on a length scale characterized by 
$1/q$, where $q = |\mathbf{q}|$.
The characteristic decay time
of $F(q;t)$ for wavevector $q$ near the peak position of the static structure factor defines a structural relaxation time, usually referred to as the $\alpha$ relaxation time, $\tau_\alpha$.
For reasons of computational efficiency, quite often one monitors the self-intermediate scattering function~\cite{HansenMcDonald} 
\begin{equation}
\label{eq:self}
F_s(q;t) = \frac{1}{N} \left< \sum_{n} e^{i\mathbf{q} \cdot [\mathbf{r}_n(t) - \mathbf{r}_n(0)]} \right>,
\end{equation}
which corresponds to the $n=m$ terms in Eq.~(\ref{eq:collective}). The characteristic decay time
of $F_s(q;t)$ for $q$ at the peak position of the static structure factor is nearly equal to $\tau_\alpha$, but easier to compute.

\begin{figure}
\begin{center}
\includegraphics[scale=0.3]{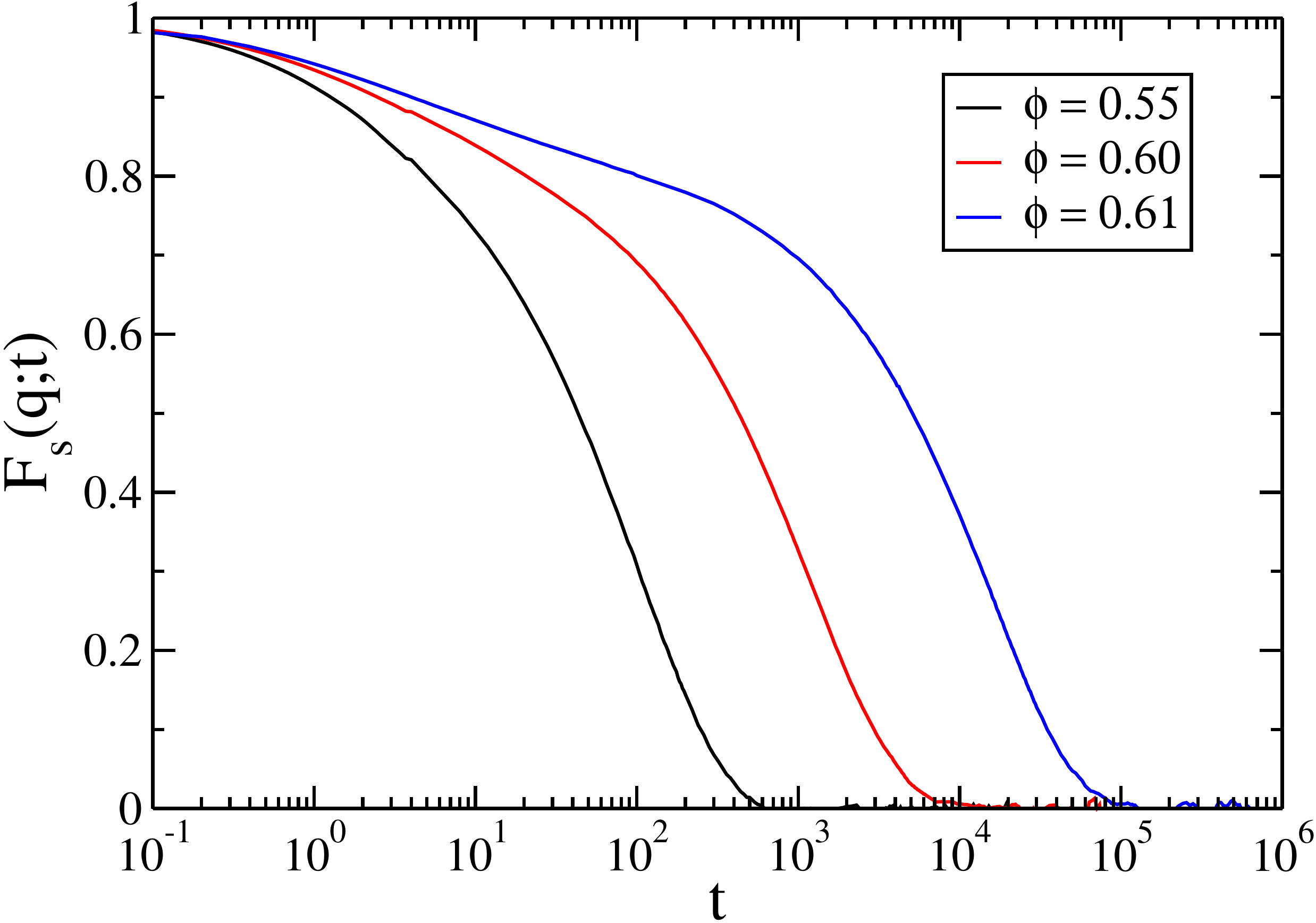}
\caption{\label{fig:pdynamics} The self-intermediate scattering function, $F_s(q;t)$, 
for three volume fractions in the vicinity of the 
apparent glass transition, for an equilibrium WCA system at a low temperature, $T=0.01$. The 
wavector $q$ is close to the position of the first peak of the static structure factor, $q=5$.
Small changes of the static structure shown in Fig. \ref{fig:pstatics} are concurrent with a 
dramatic slowing down of the dynamics.}
\end{center}
\end{figure}

In a simple liquid above the onset of glassy dynamics, it is found that $F_s(q;t)$ 
decays nearly exponentially. This is consistent with
a Gaussian distribution of particle displacements
\begin{equation}
G_s(r;t) = \frac{1}{\rho N} \left< \sum_{n} \delta[\mathbf{r} - ( \mathbf{r}_n(t) - \mathbf{r}_n(0)) ] \right>
\end{equation}
whose mean-square average increases linearly in time, \textit{i.e.} with Fickian diffusion.

Two differences occur when the liquid is supercooled or if its density is increased.  
A plateau develops in $F_s(q;t)$ that indicates that the particles are localized, as a solid, 
at intermediate time scales. The particles are said to be trapped in cages formed by their neighbors, and
they have to escape their cages for $F_s(q;t)$ to decay from the plateau.
The decay from the plateau occurs at increasingly later times upon approaching the glass transition, and 
an operational definition of the glass transition is that one is no longer willing to wait for $F_s(q;t)$ to 
decay from this plateau. The other major change is that the decay after the plateau
is no longer exponential; it is usually fitted by a stretched exponential function, $\propto \exp(-(t/\tau)^\beta)$,
where the so-called stretching exponent $\beta$ decreases with decreasing temperature. 
In Fig. 2 we show $F_s(q;t)$ for the same state points as in Fig.~\ref{fig:pstatics}.

The non-exponential decay of $F_s(q;t)$ implies that the probability of the displacements $G_s(r;t)$ is non-Gaussian. 
The non-Gaussian character of the single particle displacements was investigated in some detail. 
It was found that the particles are localized 
for an extended period of time, then make a relatively quick jump to another cage where they stay
for another extended period of time. A consequence of this hopping-like motion
is that the particles can be separated into slow and fast sub-populations. The slow particles are ones that moved less than
expected for a Gaussian distribution of displacements and the fast particles are ones that
moved more than what was expected for a Gaussian distribution of displacements. Importantly,
the slow and fast particles are also found to be spatially correlated and form increasing larger
clusters upon approaching the glass transition. These spatially heterogeneous dynamics 
are recognized as one of the hallmarks of glassy dynamics~\cite{berthier-book}. 

In the description above, we have not specified the type of microscopic dynamics giving rise to the glassy dynamics. Interestingly, it was demonstrated by direct numerical comparison that the global evolution of the relaxation dynamics, of the slow relaxation of time correlation functions, of the dynamic heterogeneity associated with spatio-temporal fluctuations of the dynamics are actually the same for Newtonian~\cite{KobAndersen}, Langevin~\cite{GleimKob}, Brownian~\cite{SzamelFlenner}, or even Monte Carlo~\cite{BerthierKob} dynamics. Physically, this implies that details of the microscopic motion at very short times do not affect the manner in which the slow dynamics proceeds at much larger times. In other words, the strong separation of timescales makes the details of the driving dynamics irrelevant at long times. This finding will play an important role when discussing the role of non-equilibrium active forces.  

\section{Driven dynamics of glasses: Rheology}

\label{sec:rheology}

As we wish to understand the behavior of dense materials driven by active forces, it is interesting to mention that glassy materials can be driven out of equilibrium by many types of forces, and `active' forces are only one particular example on which we shall focus below.

A well-known example of a driving force that is frequently applied to a dense assembly of particles is an external mechanical perturbation that can take the form of a shear flow, or a constant stress~\cite{Larson}. The obvious qualitative difference with active forces is that such mechanical perturbation is applied at large scale, rather than at the particle level but dense active particles or sheared thermal systems are two examples of non-equilibrium glassy dynamics. Before discussing the former, it is therefore interesting to learn from the latter case. 

The field of glassy materials driven by an external mechanical constraint relates to the rheology of glassy systems. Starting from an arrested glass at low temperature, the application of a constant force (such as a shear stress) may give rise to a yielding transition. Whereas the glass responds in a nearly linear manner at small applied force, the response becomes non-linear at larger applied force until a well-defined force threshold is crossed (called a ``yield stress'') above which the glass deforms plastically and undergoes microscopic relaxation. The yielding is thus a form of a solid-to-fluid transition driven by an external force of sufficient strength, which is currently under intense scrutiny~\cite{Ozawa,yieldRMP}. The response of a glassy system to an applied external force is obviously a relevant problem for researchers dealing with active glassy materials. 

Another way to mechanically drive a glass is to impose a constant rate of deformation, \textit{i.e.} a finite shear rate. This is again a useful analogy since such geometry introduces a new timescale (the shear rate) for the external driving force, in close analogy with self-propelled motion in active particle systems (where the timescale for the driving force is the persistent time of the self-propulsion).
The presence of a finite shear rate has been analyzed in great detail~\cite{BB,MRY}. The main finding is that to sustain a constant deformation rate, a dense system needs to constantly undergo plastic rearrangements, and the structure is thus never dynamically arrested. In other words, the system is always in a driven steady state where particles diffuse and the structure rearranges and there cannot be a fluid-to-solid transition since the material is permanently in a non-equilibrium driven fluid phase. 
 
Based on such analogy one could tentatively conclude that active glassy systems always flow in the presence of active forces. We shall see shortly that this intuition is not correct. 

\section{Fluid-to-solid jamming transition}

\label{sec:jamming}

The jamming transition describes a fluid-to-solid transition in the absence of any fluctuations, in particular thermal fluctuations~\cite{LiuNagelreview} or active forces. 

A clean setting to observe the jamming transition is to consider packings of soft repulsive spheres; imagine for instance green peas (without gravity). Peas are a useful image, as thermal fluctuations are clearly insufficient to drive their dynamics. The jamming transition separates a low-density regime where the assembly of peas can not sustain a shear stress and responds as a fluid, from a large-density regime where the assembly of peas responds as a solid. For repulsive spheres, the details of the jamming transition have been worked out in great detail. In particular, it is found that the emergence of rigidity corresponds to a nonequilibrium critical point, characterized by power laws and several critical exponents. In particular, the pair correlation function $g(r)$ in Eq.~(\ref{eq:pair}) develops singular behavior exactly at the jamming transition, the yield stress increases continuously from zero as a power law of the density, \textit{etc.} 

\begin{figure}
\begin{center}
\includegraphics[width=0.6 \columnwidth]{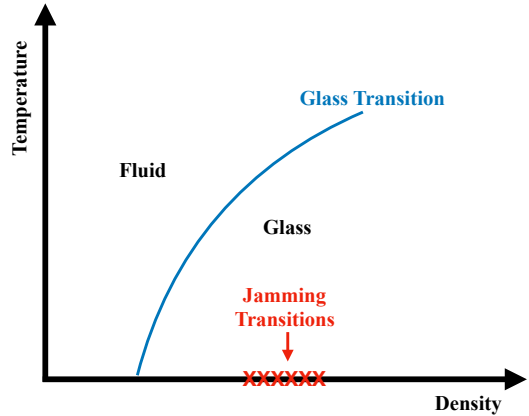}
\caption{\label{fig:rhot} 
Schematic temperature-density phase diagram for soft repulsive spheres undergoing a glass transition at finite temperature, and jamming transitions in the absence of thermal fluctuations.}
\end{center}
\end{figure}

Since the transition takes place exactly at zero temperature, it is important to realize that there is, by definition, no glassy dynamics that can be observed near the jamming transition, since the former can only emerge when particles are driven by some sort of fluctuations. A sheared assembly of such non-Brownian particles does not display glassy dynamics either. 
Another important consequence of the absence of any dynamics is that the preparation protocol of the athermal packing needs to be specified to analyze jamming~\cite{ohern,torquato}. In particular, it is found that the location of the jamming transition for a given system cannot be unique, but is instead dependent on the packing preparation. There is thus not a unique jamming transition density, but instead a line of critical protocol-dependent jamming transitions~\cite{pinaki}. 

Although glass and jamming transitions both  describe fluid-to-solid transitions, they are quite distinct physical phenomena. This is most easily realized by studying the temperature-density phase diagram of soft repulsive spheres~\cite{tom,atsushi,hugo}, see Fig.~\ref{fig:rhot}. The jamming transitions take place at $T=0$ along the density axis over a range of densities. By contrast, the glass transition with its associated glassy dynamics takes place at finite temperature, and thus the structural and dynamical signatures associated with both transitions are observed in different, non-overlapping physical regimes. 

From this phase diagram, an apparent paradox emerges: the system at $T=0$ undergoes a fluid-to-solid transition by increasing the density through the red cross at $\rho_J$ on the horizontal axis, but the system at finite $T>0$ is an amorphous glass even at densities smaller than the jamming critical density $\rho < \rho_J$. To reconcile the two opposite mechanical behaviours obtained at very low $T$ and $\rho < \rho_J$, one needs to realise that the system at any finite $T$ is actually a solid with a finite shear and bulk modulus, but the amplitude of these mechanical moduli is proportional to the thermal energy, $T$. These solids are entropic in nature. By contrast, the moduli only depend on the density $\rho$ when $\rho>\rho_J$ in the vicinity of the jamming density and thus remain solid even as $T \to 0$: they are enthalpic solids. 
    
\chapter{Dense active matter}

\label{active}

In this section we discuss the dynamics of active particle systems in the dense regime where they may undergo a dynamic arrest similar to glass transitions observed in molecular and colloidal systems, as reviewed in Sec.~\ref{sec:equilibriumgt} above.

Our presentation will follow an increasing complexity pattern. We shall start with a generic Langevin equation for a solvable glassy model~\cite{BerthierKurchan} where active forces appear as a schematic set of additional driving forces, in the spirit of the general discussion in Sec.~\ref{sec:jorgewhynot} above. 
We then move to studies of active glass transitions in computer models of self-propelled particles, before reviewing computational studies of increasingly complex theoretical models. Finally we close with a brief review of experimental studies of glassy dynamics in active materials. 

\section{Many-body mean-field model for active glass transitions}

\label{twobaths}

To model the experimental situations described above we consider the
dynamics of $N$ degrees of freedom, ${\bf  x} = \{x_i, i=1 \cdots N \}$,
representing for instance the position of grains or cells,
interacting through the Hamiltonian ${H} [{\bf x}]$, 
which is supposed to display a glass transition at thermal 
equilibrium. The driven and active materials we wish to study 
share two important characteristics. First, they dissipate energy
through internal degrees of freedom at a finite rate.
Second, energy is continously supplied either by a global external 
forcing or by the particles themselves. 
To account for these effects, we study the following equation of motion:
\be
\dot{x_i}(t) + \int_{-\infty}^t ds \gamma_d(t-s) 
\dot{x_i} (s) + \frac{\partial { H}}{\partial x_i}
+ \eta_i(t) + f_i^{a}(t) = 0,
\label{eqofmotion}
\ee
where we have included contributions from both a (white noise) equilibrium 
thermal bath 
satisfying the fluctuation-dissipation theorem, 
$\langle \eta_i(t) \eta_j(s) \rangle = 2 T  \delta(t-s) \delta_{ij}$
and from nonequilibrium, colored driving and dissipative mechanisms
represented by the active force $f_i^a(t)$
and the dissipation kernel $\gamma_d(t)$, respectively.
Equation (\ref{eqofmotion})
is a standard theoretical model for the dynamics of active colloids
and molecular motors far from equilibrium~\cite{visco,visco2,blood}. 
It also represents a minimal model to analyze the physics studied
in numerical treatments of active~\cite{mossa,mossa2} and self-propelled 
particles~\cite{marchetti}, where particles 
perform persistent random walks. It certainly misses 
some features of more complicated situations, such as 
complex alignement rules or particle anisotropy~\cite{marchetti2}.

As a first step we can choose simple functional forms 
for the colored noise and dissipation 
terms. We use a Gaussian random forcing with mean zero and 
variance $\langle f_i^a (t) f_j^a(s) \rangle = 2 F_a(t-s) \delta_{ij}$,
where
$F_a(t) = \frac{\epsilon_a}{\tau_a}
\exp(- \frac{t}{\tau_a})$,  
with $\tau_a$ the timescale of the slow forcing. 
For the dissipation we similarly choose
$\gamma_d(t) = \frac{\epsilon_d}{\tau_d} \exp(- \frac{t}{\tau_d})$, 
which defines the timescale $\tau_d$.
With these definitions, thermal equilibrium is recovered either 
when $\epsilon_d = \epsilon_a = 0$, or when the colored 
forces and friction satisfy the equilibrium FDT~\cite{kurchan}, 
$F_a(t) = T_a \gamma_d(t)$ and imposing $T_a=T$. As mentioned in Sec.~\ref{sec:jorgewhynot}, even if the slow bath obeys the fluctuation dissipation relation but with a distinct temperature, $T_d \neq T$, then the system is automatically driven out of equilibrium.  

To make the problem analytically tractable and conceptually sharper, we
perform a mean-field approximation of the glass Hamiltonian. Our goal is to have a well-understood 
equilibrium starting point to isolate the influence 
of the non-thermal forces.
We specialize our study to the spherical $p$-spin model, as a well-known 
representative microscopic model where the mean-field theory of the glass transition becomes exact~\cite{KT87}. It is defined by 
\be
{ H} = - \sum_{i_1, \cdots , i_p} J_{i_1 \cdots i_p} x_{i_1} 
\cdots x_{i_p}, 
\label{ham}
\ee
for continuous spins obeying the spherical constraint 
$\sum_i x_i^2 = N$.  
In short, our strategy is to perform a mean-field approximation to 
the interactions in the equation of motion (\ref{eqofmotion}), 
while retaining realistic forms for the sources of injection and 
dissipation. As usual when dealing with glassy dynamics, our
theoretical predictions strictly hold within the particular
context of mean-field theory, but we expect 
them to have wider physical relevance, see Ref.~\cite{BerthierBiroliRMP}
for a broad theoretical overview. Another advantage of our 
approach is that it provides precise predictions which
are then useful guides to computer simulations of more realistic 
models of active particles.
 
Because the Hamiltonian (\ref{ham}) is fully-connected, 
closed and exact equations of motion can be derived for 
the autocorrelation function
$C(t,s) = \langle x_i(t) x_i(s) \rangle$, and for the 
autoresponse function $R(t,s) = \partial \langle x_i(t) \rangle
/ \partial \eta_i(s)$:
\begin{eqnarray}
\frac{\partial C(t,s)}{\partial s} & = &
- \mu(t) C(t,s) + \int_{-\infty}^s dt' D(t,t') R(s,t') 
+ \int_{-\infty}^t dt' \Sigma(t,t') C(t',s) 
+ 2 T R(s,t), \nonumber \\
\frac{\partial R(t,s)}{\partial s} & = &
- \mu(t) R(t,s) + \int_{-\infty}^t dt' \Sigma(t,t') R(t',s) 
+ \delta(t-s), \nonumber \\
\mu(t)  & = & T + \int_{-\infty}^t dt' \left[ D(t,t') R(t,t')
+ \Sigma(t,t') C(t,t') \right] ,
\label{detailedp}
\end{eqnarray}
with the kernels
$D(t,s)  =  \frac{p}{2} C^{p-1}(t,s) + F_a(t-s)$ and 
$\Sigma(t,s) = \frac{p(p-1)}{2}
C^{p-2}(t,s) R(t,s)  + \frac{\partial \gamma_d(t-s)}{\partial s}$. 
The last equation in Eq.~(\ref{detailedp}) enforces the spherical constraint.

Technically, introducing colored friction
and noise breaks detailed balance and introduces new physical timescales
($\tau_d$ and $\tau_a$) which compete and perturb the 
equilibrium dynamics of the system. 

This situation is superficially reminiscent of 
the driven dynamics of the model studied in previous work~\cite{BBK}, 
where non-Hamiltonian driving forces were introduced to model 
an applied shear flow. The crucial difference is
the form of the driving terms, whose 
typical timescales in Ref.~\cite{BBK} were that of the dynamics itself, 
while here they relax with their own, 
fixed timescales $\tau_d$ and $\tau_a$ (cf. Eq.~(\ref{detailedp})). 
As a result, while the equilibrium glass transition was found to
disappear in the presence of power dissipation of infinitesimally small 
amplitude~\cite{BBK}, we find here that the glass transition may also survive the introduction of fluctuating forces, even of large amplitude.

\begin{figure}
\begin{center}
\includegraphics[width=8cm]{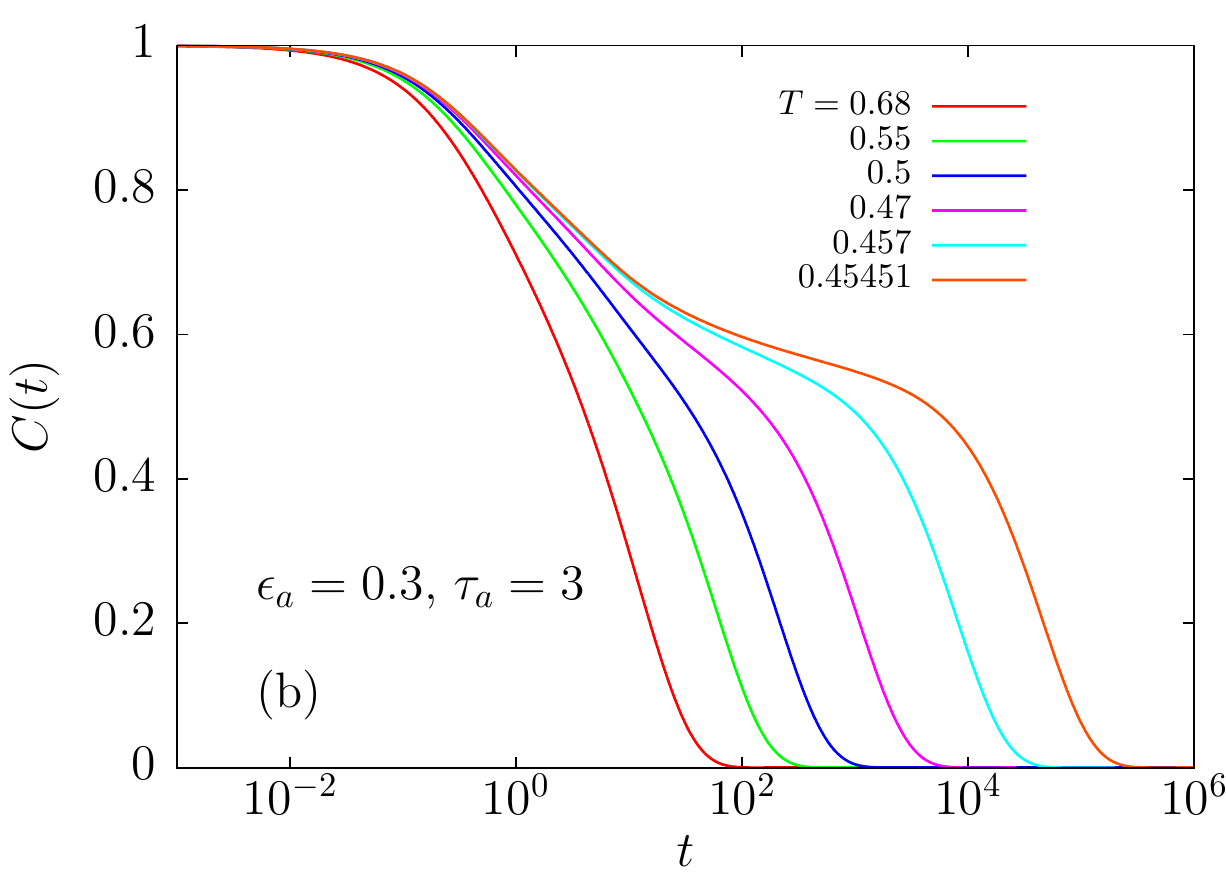}
\caption{Slow dynamics with colored noise display usual features of systems approaching a glass transition with the emergence of a two-step decay, caging, and strectched exponential relation at long times. Compare with equilibrium dynamics in Fig.~\ref{fig:pdynamics}.}
\label{corr}
\end{center}
\end{figure}



A direct analytical solution only 
exists in the trivial limit where
$\tau_d, \tau_a \to 0$, because Eq.~(\ref{eqofmotion})
reduces to a standard Langevin dynamics with 
white noise and memoryless friction. Equilibrium
is then achieved at the rescaled temperature
\be 
\bar{T} = \frac{T+\epsilon_a}{1+\epsilon_d}.
\label{barT}
\ee 

Equation (\ref{barT}) implies
in particular that response and correlation functions satisfy 
the fluctuation-dissipation theorem, ${\bar T} R(t) = - \frac{dC(t)}{dt}$, 
and that 
Eqs.~(\ref{detailedp}) reduce to 
\be 
\frac{dC(t)}{dt} + {\bar T} C(t) 
- \frac{p}{2 {\bar T}} \int_0^t dt' C^{p-1}(t-t') \frac{dC(t')}{dt'} = 0,
\label{equil}
\ee  
which is mathematically equivalent~\cite{KT87} to the
so-called $F_{p-1}$ schematic model derived in the context 
of the mode-coupling theory (MCT) of the glass transition~\cite{gotze}. 
Its solution is known in great detail, and
displays a dynamic singularity as the (rescaled) temperature 
is lowered towards the equilibrium value $T_c^{\rm eq}$. Near the dynamic 
transition, the correlation function develops a two-step decay, 
with asymptotic time dependences that follow the 
behaviour described for discontinuous (or `type B') transitions 
within MCT~\cite{gotze}. An expression for the critical temperature 
and scaling laws is obtained by performing a detailed mathematical 
analysis~\cite{gotze} of the equation derived by 
taking the long-time limit of 
Eq.~(\ref{equil}),
\be
{\bar T} C_s(t) 
+ \frac{1}{{\bar T}} \int_0^t dt' D_s(t-t') \frac{dC_s(t')}{dt'} 
- \frac{1-q}{\bar T} D_s(t) = 0,
\label{equils}
\ee 
where $D_s(t) = \frac{p}{2} C_s^{p-1}(t)$.

For general values of the timescales, we numerically solve the dynamical equations of motion to obtain the main features of the multi-dimensional phase diagram, varying $\tau_a$, $\tau_d$, $\epsilon_a$ and $\epsilon_d$.  
We illustrate our results in Fig.~\ref{corr} with the solution obtained in one limiting case where either only active driving forces  are introduced. We do not find any added complexity when both terms are simultaneously present with different amplitudes and timescales. 



The most important result is that the main features of this equilibrium 
glass transition robustly survive the introduction of a finite 
amount of non-thermal fluctuations driving the system far from thermal
equilibrium, as illustrated in Fig.~\ref{corr}. It is numerically found that time 
correlation functions display a two-step decay reminiscent 
of the equilibrium behaviour. This emerging glassy
dynamics is characterized by a relaxation time 
that diverges upon approaching a
dynamic transition, which we call a {\it nonequilibrium glass transition}. 
We find that the location of the transition 
is a continuous function of the driving mechanism. 
The numerical solutions confirm the natural expectation 
that $T_c$ increases in the presence of the 
additional colored dissipation, 
$T_c(\epsilon_d, \tau_d) > T_c^{\rm eq}$, 
while it decreases in the presence
of a colored forcing, 
$T_c(\epsilon_a, \tau_a) < T_c^{\rm eq}$. 



We can analytically rationalize the above findings, and obtain 
additional insight into the slow dynamics near nonequilibrium 
glass transitions, relying on the fact that the driving terms 
responsible for the explicit violation of detailed balance
in the dynamical equations (\ref{detailedp}) have correlations 
that vanish at long times. Thus, a 
strong scale separation occurs
when the structural relaxation is much
larger than both $\tau_d$ and $\tau_a$, i.e. 
sufficiently close to the nonequilibrium transition. 
We seek an approximate equation of motion valid 
for stationary states in the limit of
large times, $\tau_d, \tau_a \ll t$, corresponding to 
the approach and departure from the plateau: 
\begin{eqnarray}
\frac{\partial C_s(t)}{\partial t} & = & - ( \mu - I_\Sigma) C_s(t)
+ I_R D_s(t) 
+ \int_0^t dt' \Sigma_s(t-t') C_s(t')  \nonumber \\
& & + \int_0^\infty dt' \left[ 
D_s(t+t') R_s(t') + \Sigma_s(t+t') C_s(t') \right],
\label{slowcorr}  \\
\frac{\partial R_s(t)}{\partial t} & = & - (\mu -I_\Sigma) R_s(t)
+ \Sigma_s(t) I_R
+ \int_0^t dt' \Sigma_s(t-t') R_s(t'), \label{slowresp} \\
\mu & = & T + \Omega + \int_0^\infty dt' \left[ 
D_s(t') R_s(t') + \Sigma_s(t') C_s(t') \right],
\label{slowmu}
\end{eqnarray}
with 
$\Sigma_s = \frac{p(p-1)}{2} C_s^{p-2} R_s$, 
$D_s = \frac{p}{2} C_s^{p-1}$, 
and the following integrals
$I_\Sigma = \int_0^\infty dt' \Sigma_f(t')$, 
$I_R = \int_0^\infty dt' R_f(t')$, 
and $\Omega = \int_0^\infty dt' \left[ 
D_f(t') R_f(t') + \Sigma_f(t') C_f(t') \right]$.
We defined the `slow' functions $C_s(t)$ and $R_s(t)$ 
as the exact solutions of Eqs.~(\ref{slowcorr}-\ref{slowmu}), 
while the `fast' ones are defined by difference, e.g.
$C_f (t) = C(t) - C_s(t)$, and decay over time scales
that do not diverge at the transition. 
 
A crucial element of the dynamical equations (\ref{slowcorr}-\ref{slowmu}) 
governing the long-time dynamics is that the terms 
responsible for the explicit breaking of detailed balance
have disappeared. They appear very indirectly through 
time integrals over the short-time dynamical behaviour.
As an immediate  consequence, these equations can be considerably 
simplified because Eqs.~(\ref{slowcorr}, \ref{slowresp}) reduce to 
the same equation if correlation and response satisfy  
\be
R_s(t) = - \frac{X_s}{T} \frac{dC_s(t)}{dt},
\label{ansatz}
\ee
which defines the fluctuation-dissipation ratio, $X_s$, 
or equivalently an effective temperature $T_{\rm eff} = T/X_s$~\cite{teff}.
A similar ansatz holds in the long-time limit of the aging 
regime~\cite{cuku93}, i.e. in the unperturbed glass phase, and in the 
equivalent limit of vanishing shear-like forces~\cite{BBK}. 

Combining Eqs.~(\ref{slowcorr}, \ref{slowmu}, 
\ref{ansatz}), we obtain
\begin{eqnarray} 
& & ( T + \frac{p}{2} X_s q^p + \Omega - I_\Sigma 
- \frac{p}{2} q^{p-1} ) C_s(t) 
\nonumber \\
& & 
  + \frac{X_s}{T} \int_0^t dt' D_s(t-t') \frac{d C_s(t')}{dt'} 
- I_R D_s(t) = 0 
\label{final}
\end{eqnarray}
where $q$ represents the intermediate plateau height of $C(t)$. 
This equation shows that, {\it at sufficiently long times}, the dynamical equation governing 
structural relaxation is equivalent to the 
one found for equilibrium relaxation, Eqs.~(\ref{equil}, \ref{equils}),
showing that an equilibrium-like glassy dynamics emerges out 
of nonthermal forces driving the dynamics at short times.


Even then, there remain, however, two important differences 
with the equilibrium case.

First, the `coupling' parameters determining the numerical value 
of the critical temperature are `renormalized' by the microscopic details
of the driving forces through time integrals over the short-time 
dynamics, as can be seen by directly 
comparing Eqs.~(\ref{equils}) and (\ref{final}).
This explains the numerical finding 
that the location of the transition 
and the value of the plateau height
continuously depend on the details of the microscopic dynamics. Thus, an analytic 
determination of the locus of the nonequilibrium glass transition 
requires solving not only Eq.~(\ref{final}), but also 
strongly nonuniversal features of the short-time dynamics. 
By contrast, because
the long-time dynamics remains described by a discontinuous MCT 
transition, all universal features of the time correlation 
functions remain valid far from equilibrium, as
found numerically in Fig.~\ref{corr}. 

Second, while Eq.~(\ref{final}) simply involves the 
correlator $C_s(t)$, as in equilibrium, the 
response function $R_s(t)$ does not obey the  
equilibrium fluctuation-dissipation relation, but only 
an `effective' one, Eq.~(\ref{ansatz}). Note that this
predicts the existence of a nonequilibrium $T_{\rm eff}$ 
for slow degrees of freedom even for the 
stationary fluid phase, not only deep into the glass as in 
Ref.~\cite{trieste}.
This finding illustrates that nonequilibrium glass transitions 
are conceptually distinct from the equilibrium analog, 
but that a form of equilibrium-like glassy dynamics 
naturally emerges at long times.

The numerical analysis confirms the existence of effective temperatures.
The results show that $X_s$ 
behaves differently if friction ($X_s (\epsilon_d , \tau_d) > 1$),
or forcing ($X_s (\epsilon_a , \tau_a) < 1$)
dominates the physics. The former represents an unusual situation
where slow degrees of freedom appear to be colder than 
the bath~\cite{peter}. Although Eq.~(\ref{barT}) cannot 
be used to predict the actual value of $T_{\rm eff}$ in the general
case, it correctly predicts its qualitative trends and thus 
provides a simple physical argument for its variation with our 
control parameters. These two distinct cases are 
reminiscent of the distinction between adamant (`hot') 
and susceptible (`cold') molecular motors of 
Ref.~\cite{wolynes}. Tests of fluctuation-dissipation relations in active matter have also been performed in computer simulations
of active fluids~\cite{mossa,mossa2} and active glasses~\cite{Levis}.

\section{Glassy dynamics with self-propelled particles}

\label{glassyself}

\subsection{Model for self-propelled particles}

To study how `activity' interferes with `crowding', a minimal model should simultaneously capture the physics of crowding in dense particle assemblies, and those particles should be driven by active forces. 
Before modeling systems as complex as epithelial tissues or self-phoretic colloidal particles in a solvent, it is useful to learn some lessons from minimal models. In equilibrium, 
the glass transition is typically studied as a function of two control parameters, the particle density controlling crowding, and the temperature that drives the microscopic motion of the particles. 
Many investigations fix one of these control parameters 
and vary the other in order to simplify studying the phase diagram, and both directions are essentially equivalent.  

Active systems composed of self-propelled particles are characterized by two more control parameters, the persistence time of the active force and its average strength. Thus, the parameter space immediately doubles from two 
to four dimensions and the problem becomes intractable.
Since we are interested in how activity influences the glass transition, we can simplify matters by removing the effects of the thermal bath, \textit{i.e.} temperature, from the picture
(note that since there is still non-vanishing activity, we are away from any jamming transition). Additionally we can either fix
the density or fix one of the parameters that controls the active motion to examine the influence of activity on the glass transition. 

Self-propulsion in active matter model can take many forms, which are believed to yield to essentially the same behavior as far as collective behavior is concerned. A mathematically appealing minimal active matter model is a system of interacting active Ornstein-Uhlenbeck particles (AOUPs) 
\cite{Szamel2014,Maggi,Fodor2016} where
particles perform overdamped motion in a viscous fluid, thus neglecting thermal fluctuations.
The self-propulsion forces evolve in time according to the 
Ornstein-Uhlenbeck~\cite{VanKampen} stochastic process. Thus, the equation of motion for the position $\mathbf{r}_n$ of particle $n$ is
\begin{equation}\label{eom1}
\dot{\mathbf{r}}_n = \xi_0^{-1}[\mathbf{F}_n + \mathbf{f}_n],
\end{equation}
where $\mathbf{F}_n = -\sum_{m\ne n} \nabla V(r_{nm})$ is the force
originating from pairwise particle interactions and $\mathbf{f}_n$ is the
self-propulsion force acting on particle $n$. Notice that since thermal fluctuations are neglected, there is no term 
corresponding to a thermal bath in Eq.~(\ref{eom1}), and thus  
without the active force the particles would only evolve towards the closest potential energy minimum. The pair potential $V(r)$ can be any simple model for a dense fluid usually studied in the field of simple glasses, from hard spheres to WCA and Lennard-Jones potentials. 

The equation of motion for the active force $\mathbf{f}_n$ is
\begin{equation}\label{eom2}
\tau_p\dot{\mathbf{f}}_n = -\mathbf{f}_n + \boldsymbol{\eta}_n,
\end{equation} 
where $\tau_p$ is the persistence time of the self-propulsion and $\boldsymbol{\eta}_n$
is an internal Gaussian noise with zero mean and variance 
$\left< \boldsymbol{\eta}_n \boldsymbol{\eta}_m \right>_{\mathrm{noise}} = 2 \xi_0 T_{\mathrm{eff}} \mathbf{I}\delta_{nm}\delta(t-t^\prime)$;
$\mathbf{I}$ denotes the unit tensor.
The average $\left< \ldots \right>_{\mathrm{noise}}$ denotes averaging over the noise distribution. 
The parameter $T_\mathrm{eff}$, which we will refer to as the (single-particle) effective temperature, 
quantifies the noise strength and, therefore, the magnitude of the 
active forces.

\subsection{Lessons from the dilute limit}

Before discussing dense systems it is useful to consider the dynamics of a single particle evolving
according to Eqs.~(\ref{eom1}-\ref{eom2})~\cite{Szamel2014}. The mean squared displacement of a single AOUP can be calculated as  
\begin{equation}
\left< \delta r^2(t) \right> 
= \frac{6 T_\mathrm{eff}}{\xi_0} \left[\tau_p \left(e^{-t/\tau_p} -1 \right) + t\right],
\end{equation}
which exhibits typical features of a persistent random walk. Indeed, 
for $t \ll \tau_p$ we can expand the exponential, $\left< \delta r^2(t) \right> 
\approx (3 T_\mathrm{eff} \tau_p/\xi_0) t^2$
and the motion is ballistic. For $t \gg \tau_p$ the exponential can be neglected, $\left< \delta r^2(t) \right> 
\approx (6 T_\mathrm{eff} /\xi_0) t$
and the motion is diffusive with a diffusion coefficient $D_0 = T_\mathrm{eff}/\xi_0$.
Here we see the origin of the name effective temperature: $T_\mathrm{eff}$ plays the same role as the
equilibrium temperature $T$ in the expression for the long-time diffusion coefficient of an isolated particle. 
Importantly, systems with the same effective temperature will have the same long time
diffusion coefficient in the absence of interactions. This makes $T_\mathrm{eff}$ a useful parameter to determine 
how the long-time dynamics changes upon approaching the glass transition. The persistence time $\tau_p$ 
gives the timescale for the transition from ballistic to diffusive motion for an isolated particle. 

After the introduction of the effective temperature $T_\mathrm{eff}$, it is natural to ask 
whether this parameter has other properties of the temperature  
in equilibrium passive systems. This question can be asked several ways. For example, one could ask
whether there is a linear response relation involving a single AOUP in which the equilibrium temperature 
$T$ is replaced by $T_{\mathrm{eff}}$? One could also ask whether 
the familiar Gibbs-Boltzmann distribution is recovered when a single AOUP is placed in an external potential, with the equilibrium temperature
replaced by $T_{\mathrm{eff}}$.

The answers to the above questions vary~\cite{Szamel2014}. It is possible to come up with a single
particle linear response problem in which, in the small frequency limit, the response and correlation
functions are related by $T_{\mathrm{eff}}$. Also, one can show that a single AOUP in a linear 
potential with a lower wall (the sedimentation problem), the probability
distribution has the Gibbs-Boltzmann form with the equilibrium temperature replaced by  $T_{\mathrm{eff}}$.
However, one can also show that the probability distribution of a single AOUP placed in a harmonic potential
has a Gaussian form, but the parameter that replaces the equilibrium temperature is in fact a function
of both $T_{\mathrm{eff}}$ and the persistence time $\tau_p$. These results suggest that, in general, $T_{\mathrm{eff}}$ does not always play the same role as the temperature in equilibrium systems. We note 
that in systems of interacting AOUPs, other temperature-like parameters could be defined~\cite{Levis,TeffEPL}. These temperatures will be influenced by the single particle effective temperature $T_{\mathrm{eff}}$, persistence time $\tau_p$
and the interparticle interactions. 

This, however,  does not preclude using the single particle effective temperature 
$T_{\mathrm{eff}}$ as a  {\rm control parameter} for dense active suspensions (with no mention of its thermodynamic interpretation) 
since it can still tell us how much the interactions slow down the long-time dynamics. Therefore, the minimal model of active glassy dynamics 
involves the single particle effective temperature $T_{\mathrm{eff}}$, the persistence time $\tau_p$ 
and the number density as control parameters. This set of control parameters allows us to investigate
the influence of {\it being driven by non-equilibrium active forces} on the glassy dynamics.

In the limit of vanishing persistence time, the equations of motion (\ref{eom1}-\ref{eom2}) reduce to the equilibrium dynamics of an
overdamped Brownian system at the temperature equal to the effective temperature. Thus, the departure from equilibrium is quantified by the persistence time, and increasing the persistence time drives the system further away from equilibrium.
For the sake of brevity, we will sometimes use the phrases increasing/adding
activity to indicate increasing the persistence time.  Note, finally, that for the hard sphere interaction, the absolute value of $T_{\rm eff}$ does not compete with any energy scale, and the system is left with only two control parameters, density and persistence time. 

\subsection{Many-body physics at large density}

\subsubsection{Basic observation: Nonequilibrium glass transition}

Armed with a simple model of active particles, we can now examine if the glass
transition exists and how it evolves with changing $T_{\mathrm{eff}}$, 
the persistence time and the density. Initial studies of hard and soft spheres suggested
that adding activity does not destroy the glass transition, but rather pushes the 
transition to a higher density, in the case of hard spheres~\cite{Ni}, or to a lower temperature 
at constant density, in the case of soft spheres~\cite{Mandal2016}. 

It may appear logical that a driven system has a delayed glass transition, as compared to its equilibrium counterpart. We will show below a counterexample that proves that intuition incorrect. We recall that another incorrect intuition could be drawn from the analogy with driven glassy systems discussed in Sec.~\ref{sec:rheology} above, where we showed that a glass driven with a given deformation rate does not possess a glass transition and is always in a non-equilibrium steady state.
Simulations and theoretical analysis for self-propelled particles show that the local (as opposed to global mechanical deformation) nature of the driving in fact qualitatively changes the picture. A self-propelled particle system does undergo dynamic arrest to an amorphous glass that we call a {\it nonequilibrium glass transition}. This expression makes clear the distinction with the equilibrium glass transition that is observed in dense particle systems driven by thermal fluctuations, as described in Sec.~\ref{sec:equilibriumgt}. 

\begin{figure}
\begin{center}
\includegraphics[width=7cm]{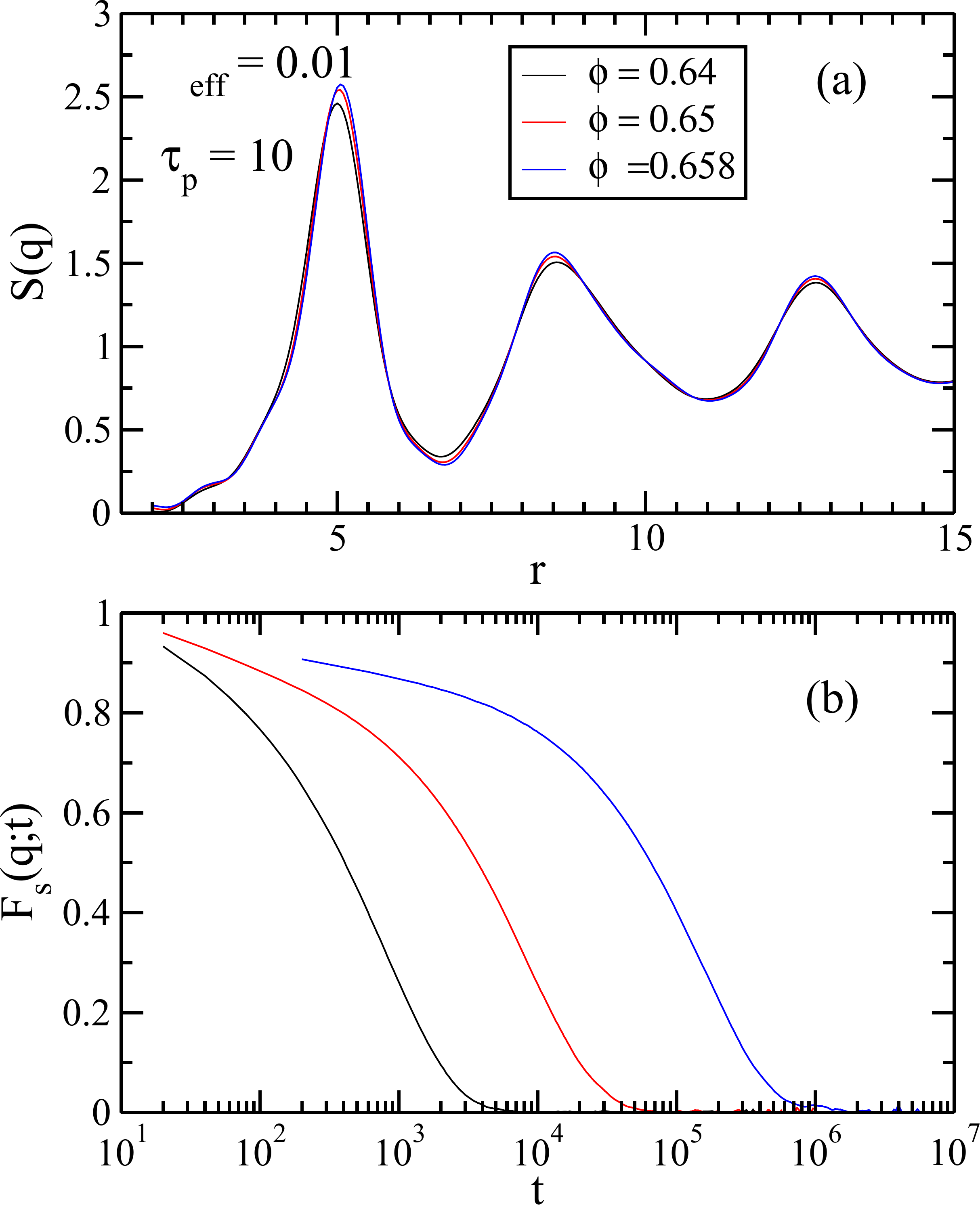}
\end{center}
\caption{\label{fig:noneq} 
Dense systems of self-propelled particles undergo a nonequilibrium glass transition as density is increased at constant effective temperature. Small variations of the steady state structure factor $S(q)$ shown in panel (a) are concurrent with dramatic slowing down of the relaxation of the self-intermediate scattering function $F_s(q,t)$ shown in panel (b).}
\end{figure}

To elucidate the role of the activity we investigate the structure and dynamics of systems of AOUPs
with the WCA interaction. Since there are three control parameters, we fix the effective temperature
at two illustrative values and then investigate the density and persistence time dependence of 
the structure and dynamics at each $T_{\mathrm{eff}}$. The two values of the effective temperature
correspond to two limits of the WCA interaction. At the higher temperature, $T_{\mathrm{eff}}=1.0$, the 
particles are able to explore a significant range of the repulsive part of the pair interaction. At the lower temperature, $T_{\mathrm{eff}}=0.01$, 
the particles do not penetrate the repulsive wall of the potential and they should behave effectively almost like hard
spheres. Thus, with these two values of $T_{\mathrm{eff}}$ we hope to analyze the behavior of a broad class of representative model systems, from models for dense liquids to dense assemblies of repulsive colloids and grains.

The central outcome of most numerical studies of dense systems with self-propulsion is that, as the strength of the self-propulsion is decreased, \textit{i.e.} as the effective temperature is decreased, or as the `crowding', \textit{i.e.} density is increased, the material undergoes a form of dynamic arrest characterized by a phenomenology very similar to observations in equilibrium systems driven by thermal fluctuations~\cite{Ni,Berthier2014}. We demonstrate these central observations in Fig.~\ref{fig:noneq} where we show the modest evolution of the pair structure of the AOUP model, which accompanies the dramatic slowing down of the dynamics and clear dynamic heterogeneity. In fact, to an unexpert eye, the data in Fig.~\ref{fig:noneq} could very well be taken as classic signatures of the glassy dynamics usually observed in equilibrium liquids, but they are reported here for a driven active system of self-propelled particles. 

\subsubsection{Nonequilibrium structure of active fluid}

Let us now turn to a more detailed description of the physics associated with nonequilibrium glassy dynamics of active particle systems. We argued in the previous paragraph that active materials display all classic features of the dynamics observed in equilibrium fluids approaching their glass transitions. Thus, our goal here will be to emphasize the new features and difficulties that are specific to active systems.  

\begin{figure}
\includegraphics[width=6.5cm]{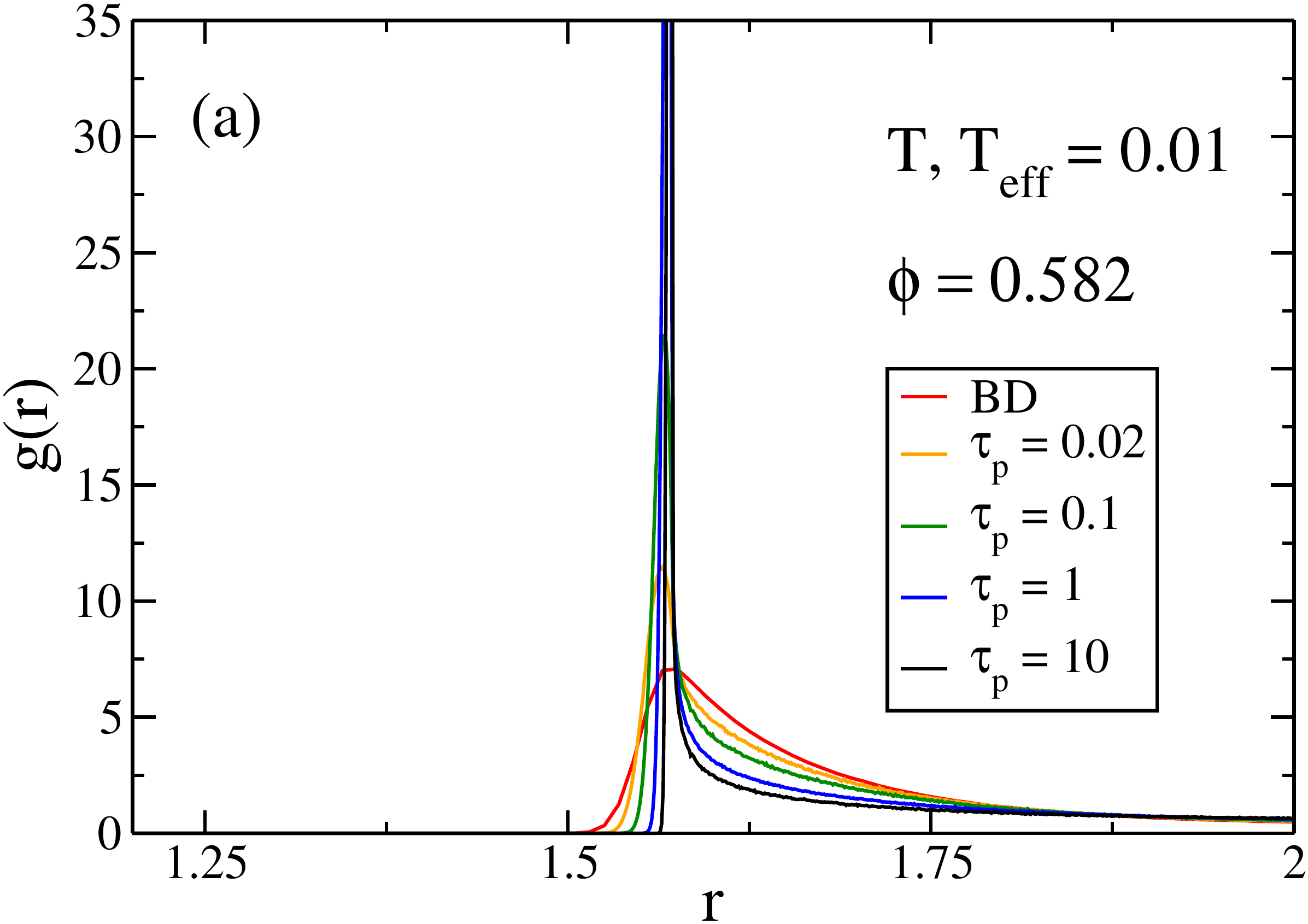}
\includegraphics[width=6.5cm]{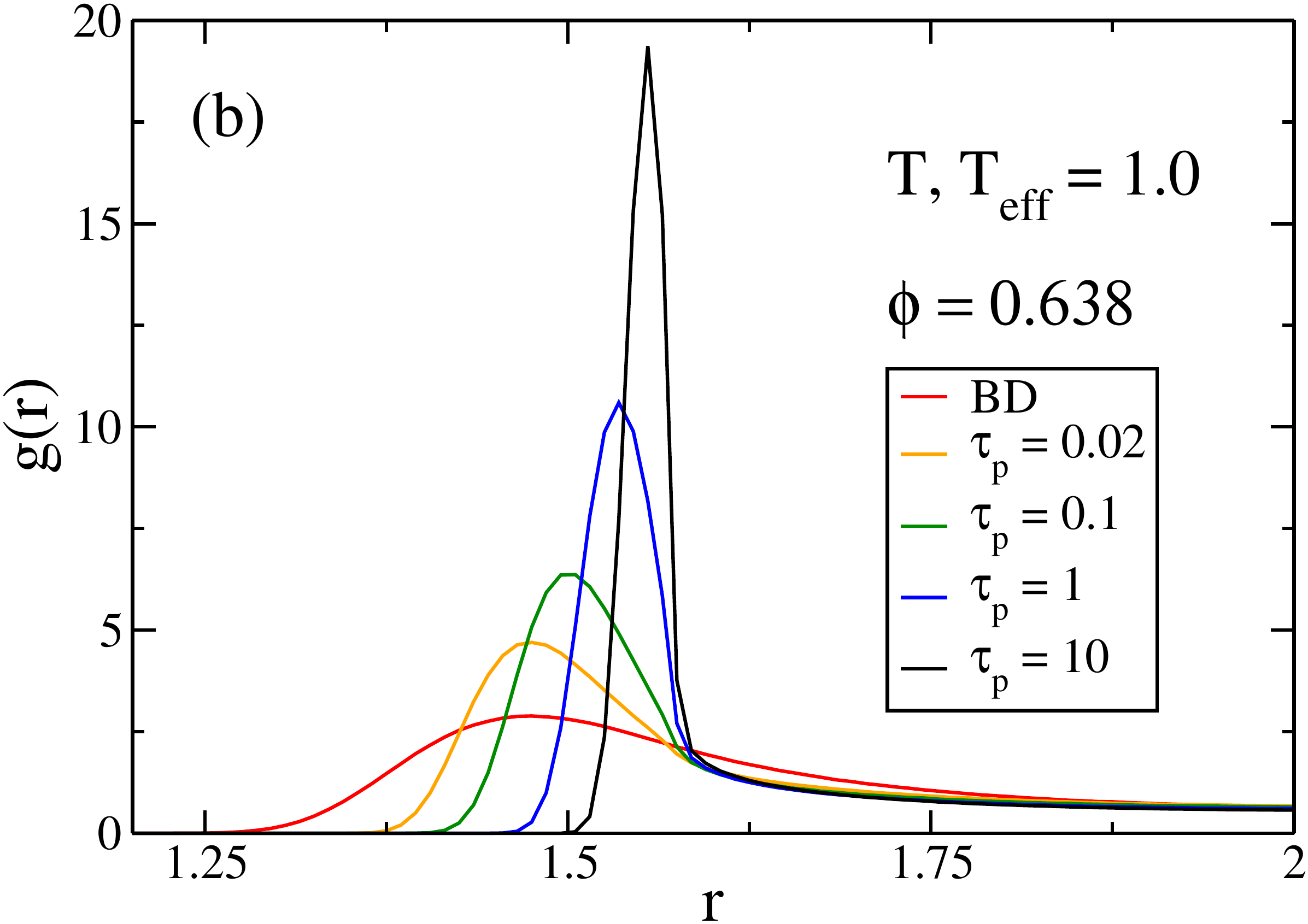}
\caption{\label{fig:agr} The steady-state pair correlation function $g(r)$ for an AOUP system with WCA interactions 
at (a) low effective temperature, $T_\mathrm{eff}=0.01$, and 
(b) high effective temperature, $T_\mathrm{eff}=1.0$. 
In the former case, particles almost never overlap but the self-propulsion leads to an effective attractive interaction that makes the particles ``sticky''. In the latter case there is some interpenetration and some effective attractive interaction but the main consequence of the self-propulsion is the increase of the effective particle radius with $\tau_p$. 
} 
\end{figure}

We start with a description of the structure of the active system approaching the glass transition. For a fixed value of the persistence time, we have shown that the pair structure evolves very little as the glass transition is approached. It is, however, interesting to ask; how does the pair structure evolve as the system increasingly departs from equilibrium with increasing the persistence time?
In Fig.~\ref{fig:pstatics} we fix the value of the effective temperature, and show how the pair correlation function $g(r)$ changes as $\tau_p$ increases. Recall that as $\tau_p \to 0$, the system is at thermal equilibrium at a temperature $T_{\rm eff}$. 

We observe that the increasing activity has a profound
influence on the pair structure of nearest neighbors. The first peak of the pair correlation
function increases very rapidly with increasing persistence time, it reaches very large values and becomes extremely narrow at the lower effective temperature. We believe that equilibrium 
systems with similar short distance structure would be totally arrested. 

More in detail, we observe that at low $T_{\rm eff}$, the position of the first peak of $g(r)$ remains at the same distance $r$ corresponding to the cutoff of the WCA potential, and thus to the radius of the equivalent hard sphere system. The growing peak amplitude can be interpreted as an effective short-range attraction resulting from the competition between the repulsive interaction and the self-propulsion. This effective adhesion has been discussed in the context of motility-induced phase separation and cluster formation in self-propelled particles~\cite{Theurkauff,ginotPRX,tailleur}. 

For the system at higher $T_{\rm eff}$, the growth of the peak amplitude is observed but is less pronounced than for the hard sphere limit. This reflects a more subtle change in the effective interaction between particles. Perhaps the more striking observation is that the peak position is highly sensitive to the persistence time and shifts to larger distances as $\tau_p$ increases. Physically, this means that the effective radius of the particles is actually increasing as the persistence time grows, suggesting an increasing crowding of the particles. 

We will discuss below the dynamical phenomena that complement these observations. For equilibrium systems, a very accurate liquid state theory was developed decades ago to predict the fluid structure starting from the pair interaction~\cite{HansenMcDonald}. There exists at present no such theory for active matter, but we clearly observe that such theory should take into account the details of the self-propulsion mechanism. The non-equilibrium nature of the self-propulsion dynamics implies that the sole knowledge of the interaction potential between particles is not enough to predict the structure of the non-equilibrium fluid. 

\subsubsection{A purely nonequilibrium object: velocity correlations} 

For equilibrium systems, the static structure is characterized by either $g(r)$ or by $S(q)$, which are the most
important structural quantities. In fact, almost all theories of glassy dynamics use the pair structure
as the only static input. 

An important development originating from the theoretical description of
dense active systems is the discovery that an additional correlation function appears in active systems that has no equilibrium analog~\cite{glassline,Szamel2015,Flenner2016,Bertin}. 
This function quantifies correlations of the 
velocities of the individual particles. The velocity of overdamped AOUP $i$ is equal to
$\xi_0^{-1}\left(\mathbf{f}_i+\mathbf{F}_i\right)$, recall Eq.~(\ref{eom1}), and the velocity correlation function
in Fourier space is defined as 
\begin{equation}\label{omegadef}
\omega_{||}(q) = \hat{\mathbf{q}} \cdot 
\left< \sum_{i,j=1}^{N}\left(\mathbf{f}_i+\mathbf{F}_i\right)\left(\mathbf{f}_j+\mathbf{F}_j\right)
e^{-i\mathbf{q}(\mathbf{r}_i-\mathbf{r}_j)}\right>\cdot\hat{\mathbf{q}},
\end{equation} 
with $\hat{\mathbf{q}} = \mathbf{q} / | \mathbf{q} |$. We note that for a binary mixture
there would be three different partial correlation functions of the overdamped
velocities. 

\begin{figure}
\begin{center}
\includegraphics[scale=0.35]{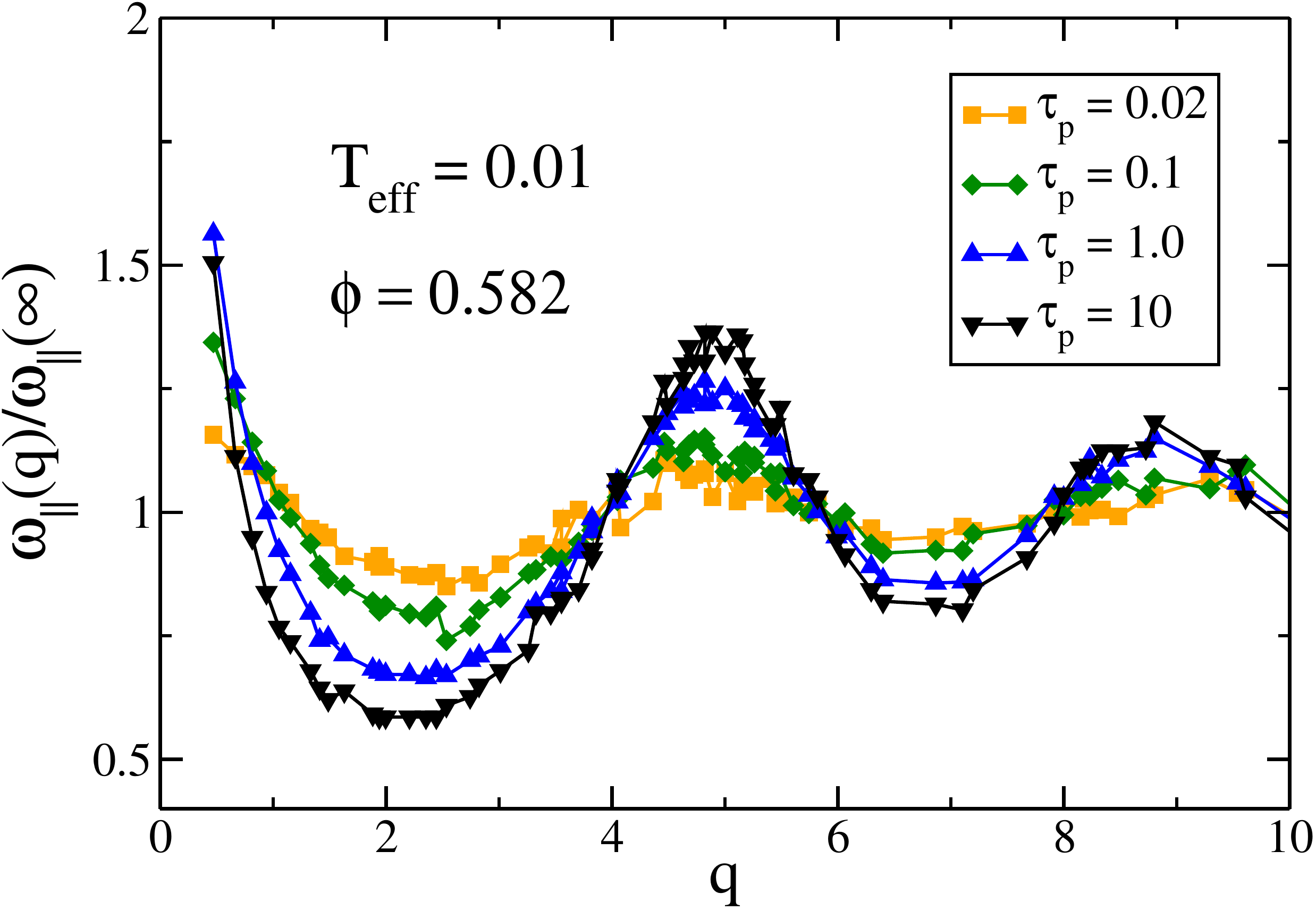}
\end{center}
\caption{\label{fig:aomega} The steady-state equal-time velocity correlations 
$\omega_\parallel(q)$ for an AOUP system with WCA interactions 
at a low effective temperature, $T_\mathrm{eff}=0.01$. Increasing range of the velocity correlations with increasing $\tau_p$ is signaled by the growth of the small wavevector peak of $\omega_\parallel(q)$. Developing local structure of the velocity correlations is evident from the growth of the amplitude of the oscillations of $\omega_\parallel(q)$. 
}
\end{figure}

In the limit of vanishing persistence time the correlation function in  Eq.~(\ref{omegadef}) becomes
trivial, \textit{i.e.} wavevector independent, because positions and velocities are independent quantities at thermal equilibrium. 
For finite persistence times it has a non trivial wavevector
dependence, as shown in Fig.~\ref{fig:aomega}. In addition to the large $q$ oscillations that imply local velocity correlations reflecting the local structure of the dense liquid, there is a clear upturn at low-$q$ that can be used to define a finite correlation length for velocity correlations.  

Physically, the non-trivial character of the velocity correlation function implies that a snapshot of short-time displacement fields is likely to reveal large-scale correlations that are purely due to the non-equilibrium nature of the active particle system. These spatial correlations represent a non-trivial form of collective motion. We note that these correlations exist even in the dense, but non-glassy, active liquid and are thus not specifically connected to the glassy dynamics itself. Numerical measurements indicate that the temperature dependence of the velocity correlation function is relatively modest, suggesting that correlations already present in the active fluid survive but do not change in any remarkable way as the non-equilibrium glass transition approaches. 

The theoretical importance of the velocity correlations (\ref{omegadef}) is twofold. First, these correlation functions enter into the exact description of the short time dependence of various
correlation functions. Second, they also enter into approximate theories of the long-time dynamics of active glassy systems. 

\subsubsection{How does activity change the slow dynamics?}

We now turn to the dynamics. We note that, quite surprisingly, in some cases the evolution of the relaxation time for a fixed $T_{\mathrm{eff}}$ does not change monotonically with $\tau_p$~\cite{Szamel2015}. For small $\tau_p$ 
the relaxation time may initially decrease and then increase
with increasing $\tau_p$. This finding demonstrates that the activity can alter
the glassy dynamics in rather subtle, unexpected ways. This non-monotonic 
behavior of the relaxation time is not mirrored in structural 
quantities such as $g(r)$ and $S(q)$, since for instance  
the height of the first peak of $g(r)$ increases monotonically with persistence
time, even though the relaxation time does not. 

An enhancement of the structure, as given by the increase of the peak height of $g(r)$ and a decrease of the relaxation time is contrary to what is expected from studies of equilibrium glassy liquids. A well-studied theory for passive liquids that connects the 
liquid structure with dynamics is the mode-coupling theory~\cite{gotze}, which has only the
static structure factor as input. While the mode-coupling theory for the glass
transition is not an exact description for this transition, it describes reasonably well the initial part of the slowing down of the dynamics and it provides microscopic physical insights.  

To gain some insight into why active systems may have a non-monotonic evolution
of the relaxation time with the persistence time, a mode-coupling-like theory 
for active systems was developed~\cite{Szamel2015,Szamel2016}. It was shown that if the theory incorporated the nontrivial character of the velocity correlations, the theory could indeed predict a non-monotonic evolution of the relaxation time with increased persistence time. There is a minimum of the relaxation time with increasing persistence time and the relaxation time begins to increase
again with increasing persistence time. The additional velocity correlations are, therefore, an important component of the slow dynamics of dense active systems.

\begin{figure}
\begin{center}
\includegraphics[scale=0.35]{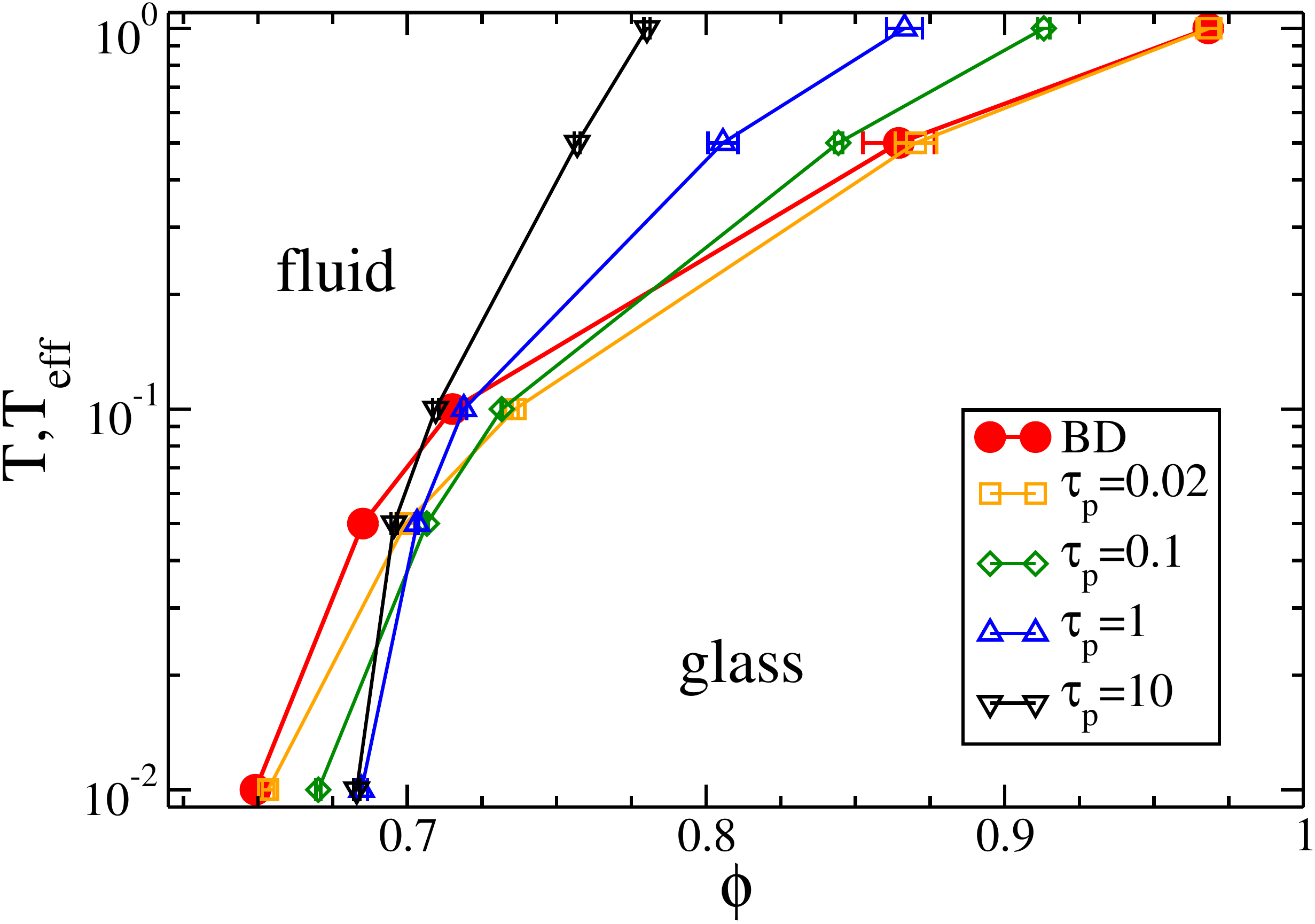}
\end{center}
\caption{\label{fig:glasspd} Evolution of the fluid-glass phase diagram with the persistence time of the self-propulsion.  With increasing persistence time the effective glass transition line (determined by fitting the volume fraction dependence of the relaxation time) shifts towards smaller volume fractions at higher effective temperatures and towards larger volume
fractions at low effective temperatures, so departure from equilibrium can either promote or suppress the glassy dynamics. Filled symbols are Brownian dynamics (BD) simulations. 
} 
\end{figure}

Next, we focus on the glass transition itself~\cite{Flenner2016,glassline}. For a fixed value of the persistence time, we find that the relaxation time increases when the effective temperature is decreased and/or when the density increases, just as for dense equilibrium fluids. This simply means that even for active systems, the glass transition results from a competition between active forces that make the particles move, and crowding that tends to arrest them. 

To analyze the increase of the relaxation time of the system and to obtain the fluid-glass phase diagram we use an empirical fitting form, $\tau_\alpha \sim \tau_0 \exp (B / (\phi_0 - \phi))$ to calculate a critical density for the glass transition, $\phi_0(T_{\rm eff}, \tau_p)$, which depends on the other two parameters of the model. The evolution of the glass transition lines are reported in Fig.~\ref{fig:glasspd}, in a (temperature, density) phase diagram. For a given value of the persistence time, the phase diagram offers two phases, the fluid at low density and high temperature, and the glass at large density and low temperature. The `BD' line is obtained from simulations performed with Brownian dynamics, i.e. in the equilibrium $\tau_p \to 0$ limit, and it corresponds to the equilibrium glass transition. All other lines correspond to non-equilibrium glass transition lines. 

The influence of a finite persistence time on the glass transition is obvious. These data confirm that at low $T_{\rm eff}$ an increase of the persistence time shifts the glass transition towards large densities, whereas the opposite effect is observed for larger $T_{\rm eff}$, with a complex behavior at intermediate $T_{\rm eff}$ values. These non-trivial dependencies show that departing from equilibrium can either promote or depress glassy dynamics, and that it is difficult to form a physical intuition, even for very simplistic models such as AOUPs. 

\section{More complex models of dense active matter}

In this section we introduce a number of more complex computational models of dense active matter.
These models were proposed to incorporate specific features of activity encountered in laboratory active matter systems. 

\subsection{More models for self-propelled particles}


Perhaps the most popular model for active particulate systems is that of active Brownian particles~\cite{tenHagen,FilyMarchetti}. 
It models active matter as consisting of particles that move via a combination of active, directed motion
and Brownian motion. The particles are endowed with an axis of symmetry. They move systematically along
this axis with a constant velocity $v_0$. The equation of motion for the 
positions takes the same form as in Eq.~(\ref{eom1}), but the active force now takes the form  $\xi_0 v_0 {\bf n}_i$, where the unit vector ${\bf n}_i$ indicates the direction of self-propulsion. 
The direction of the axis of symmetry moves via rotational diffusion with diffusion coefficient $D_R$. 
In two dimensions, that diffusion is described by a single angle that evolves via a simple Langevin equation of the form $\dot{\theta} = \sqrt{2 D_R} \eta$, where $\eta$ is a Gaussian white noise.
In addition, the particles may also be subjected to a random Brownian force and instantaneous friction characterized by temperature $T$ and friction coefficient $\xi_0$. The particles interact via spherically symmetric interaction potential $V(r)$. 
In the original version of the model, the resulting
translational diffusion coefficient due to these random forces, $D_T=T/\xi_0$ and the rotational diffusion coefficient were constrained to follow 
the relation imposed by hydrodynamic considerations~\cite{tenHagen}, 
$D_R= 3D_T/\sigma^2$, 
where $\sigma$ is the particle diameter. In several studies this relation has been relaxed~\cite{Ni} and
both $D_R$ and $D_T$ were treated as independent parameters, which is equivalent to using the persistence time of the AOUP particle  as a free parameter of the model.

The active Brownian particles 
model is intended to represent active colloids. It has been used in particular in many studies of
motility-induced phase separation. It has also been used to study the influence of the activity on
the glassy dynamics. In a simulational study of active Brownian hard spheres, 
Ni \textit{et al.}~\cite{Ni} showed that when the magnitude of the systematic velocity is increased while all the other parameters are kept constant, the apparent glass transition volume fraction moves towards larger values. 
Ni \textit{et al.} noted that the faster dynamics was accompanied by decreasing height of the first peak
of the steady state structure factor. This finding qualitatively agrees with the results of the investigation of the glassy phase diagram described in the previous section, where this corresponds to low $T_{\rm eff}$ values for the AOUP model. In another study, Fily {\it et al.}~\cite{Fily2014} 
used the active Brownian particle model with a soft repulsive potential to map out the density-temperature phase diagram of the model. They also reported a `frozen' phase at low activity and large density, which in our view should be interpreted as a glass, but the slow glassy dynamics on the approach to this arrested glass phase was not analyzed in detail. We expect that it should present the same phenomenology as the AOUP models shown above.

\subsection{Aligning interactions}

The field of active matter was largely born from the quest to describe and understand theoretically the physics of animal flocks. The Vicsek model~\cite{Vicsek} was conceived to capture the competition between the natural tendency for animals to align the direction of their self-propulsion and an external noise. While it is unclear whether all self-propelled particles types (such as cells and colloidal particles) truly possess the same tendency to alignment, the existence of {\it implicit} aligning interactions was demonstrated for some active materials, such as vibrated polar disks~\cite{Deseigne}.

A computational model was proposed in Refs.~\cite{model-olivier,model-olivier2} to 
describe a system of vibrated polar disks first studied in Ref.~\cite{Deseigne}. 
In this computational model, the motion of the particles is not overdamped. The particles are endowed with a polarity represented by a unit vector. Thus, the instantaneous state of a given particle is represented by its 
position, velocity and polarity. The self-propelling force of constant magnitude acts in the direction
of the polarity. In turn, there is a torque acting on the velocity, which tends to align it with the polarity.
There might also be two stochastic torques that randomly rotate the velocity and the polarity vectors.
Finally, the particles interact via a spherically symmetric interaction potential. This is quite a complicated model with many adjustable parameters that leads to a huge parameter space. However, since it was first
proposed to describe features observed in a specific experimental work, 
the parameters where adjusted to best reproduce the results of that specific experiment. It was shown that single particle, binary and collective properties of the experimental system can indeed be reproduced numerically.
Both the experimental and the model have also then been studied at larger density when particles form an active crystalline phase~\cite{briand}. 
Computational and experimental studies of the glassy phase of a binary mixture of the same model are currently in progress and preliminary results suggest that an active glassy phase is indeed found, whose properties will hopefully be analyzed in more detail in future work. We note that the model of Refs.~\cite{model-olivier,model-olivier2} could be thought of as an under-damped version of a model analyzed in Ref.~\cite{marchetti}. The latter model also exhibits implicit aligning interactions. The authors of Ref.~\cite{marchetti} identified a `jammed' phase that in our view is an arrested glass phase. Again the transition between the fluid and arrested phases was not characterized in any detail, and this would be a worthwhile research effort.  

In an effort to describe the collective motion observed in dense epithelial tissues, Sepulveda {\it et al.}~\cite{hakim} proposed a computational model where particles interacting with a rather complex pairwise interaction are self-propelled with a finite persistence time and are subject to short-range aligning interactions between the direction of the self-propulsion. Again, the parameter space of the model is impressive, but the many parameters of the model were adjusted to reproduce a specific set of experimental observations. In some later versions of the model, friction to a substrate and additional ingredients were added to the model~\cite{Garcia2015}. Finally, we note yet another model with aligning interactions introduced in Ref.~\cite{cerbino}.
It would be interesting to try and simplify such models in order to address more specifically the physical question of how aligning interactions between self-propulsion directions may affect, and perhaps change qualitatively the glassy dynamics obtained in the absence of aligning interactions reviewed above. 

\subsection{Modeling cell dynamics: division and death}

One of the common phenomena in biological active matter is cell division and death. A combination of these
processes can lead to an unstable system with cells eventually dying out, or instead a growing tissue that expands and invades space. If cell death and division are instead statistically balanced, a driven steady state can be reached. Quite importantly, for the present article, it was shown by Matoz-Fernandez \textit{et al.}~\cite{Matoz} that cell division and death can strongly influence the glassy behavior. 
In this study, a particle-based model of two dimensional ephitelial tissues was investigated. 
The particles interacted via a combination of a short-range repulsive and a longer-range attractive harmonic 
potentials. The activity was modeled as a combination of a cell death process, in which particles representing
cells were randomly removed from the system and a cell division process, in which a new (daughter) cell
was added on top of an existing (mother) cell, with a probability depending on the number of neighboring
cells in contact with the mother cell. 

A rich non-equilibrium phase diagram with gas-like, gel-like and dense 
confluent phases was found. A remarkable result is that in the dense, confluent phase any positive rate of cell death and division always fluidizes the system and prevents any amorphous solidification. Physically the reason is that any such event reorganizes the system locally, and thus at long times any location in the system has eventually reorganized with probability unity, completely reshuffling the structure; the dynamics is not arrested. 
In contrast, a system without cell death and division but endowed with activity modeled
similarly to that present in the active Brownian particles model, was found to exhibit classic features of
glassy dynamics upon decreasing the magnitude of the velocity, as expected by analogy with the type of minimal active model discussed above.

In the opposite limit where death is not compensated by cell division the density of cells would increase exponentially with time in a confined volume.  Here the interesting setting is when open boundary conditions are present, since from just a few cells that can divide a large tissue/colony can expand. 
This was numerically modeled in Ref.~\cite{Malmi} using an appropriate pairwise interaction and a dynamics uniquely ruled by the stochastic rules for particle division. In agreement with the steady state study of  Matoz-Fernandez \textit{et al.},  Malmi-Kakkada \textit{et al.} \cite{Malmi} conclude that cell division also leads to a complete reshuffling of the growing colony at long times, suggesting that cell division rate directly controls the onset of glassy dynamics. However, they surprisingly do not observe any specific glassy feature even when the division rate is small. It would be interesting to specifically revisit such a model in order to analyze in more detail the microscopic mechanisms responsible for tissue fluidization~\cite{ranft,prost}.
 
\subsection{Vertex models for tissue morphology}

Finally, let us briefly discuss two models that, to different degrees, are not particle based.
These models belong to the category of vertex-like models that are very popular among researchers
focusing on modeling real confluent biological tissues. The first model is the so-called Voronoi 
model~\cite{biPRX}.
In this model, the cells are modeled as Voronoi volumes defined by their neighbors and the degrees of freedom
are the Voronoi cell centers. However, the energy expression is that of the standard vertex model~\cite{vertex}, 
where the energy is given as the sum of quadratic departures of the area and the perimeter from
their preferred values. In this model the forces act on the Voronoi cell
centers. The second model is the standard vertex model~\cite{vertex}, in which the degrees of freedom (on which the
forces and thermal noise are acting) are the positions of the vertices of each cell. The same energy expression 
is used as in the Voronoi model.  

Vertex models are interesting models because they reflect more faithfully the geometric structure of dense confluent tissues. A remarkable result is that the vertex model may undergo a jamming transition in the absence of driving that is purely controlled by the competition between surface and bulk terms in the energy functions~\cite{biNature}. Therefore, as the average shape of the cells evolves the mechanical response of the system changes from a fluid to a solid response, in very much the same way a dense packing of soft particles undergoes a jamming transition as the density is increased, as discussed in Sec.~\ref{sec:jamming}. 

The properties of these models have also been studied in the presence of either thermal forces (\textit{i.e.} in equilibrium), or in the presence of self-propulsion with a finite persistence time~\cite{biPRX,daniel,lisa}, and very recently when cell division and death also take place~\cite{michael}. For a given persistence time (that can be zero), a transition between a fluid and an arrested solid state is observed, with a growing relaxation time and, again, the phenomenology associated with a nonequilibrium glass transition suggesting that a phase diagram for vertex models in a plane comprising activity and parameter shape should qualitatively resemble the sketch in Fig.~\ref{fig:rhot} for soft spheres. Further work should clarify the details of both the glass and the jamming transition in the broad family of vertex and Voronoi models. 

\section{Glassy dynamics in experimental active matter}

In the last decade it has been realized that many, if not all, of the phenomena associated with glassy dynamics
could also be observed in dense active matter systems. For the purpose of this article, the term 
active matter encompasses a variety of different materials~\cite{review1,review2,review3}. They range from living tissues to systems of active 
colloidal particles to macroscopic granular objects driven by mechanical perturbations. The differences 
between these very diverse systems have consequences for the phenomena that can be observed and for the
details of the corresponding experiments. 

Let us start with some specific experiments on cells and tissues. Typically, in these systems cells are proliferating and sometimes
also dying, with the overall cell density being a non-trivial function of time during the experiment. Since 
the dynamics of dense systems is very sensitive to their density, the fact that the number density is changing
imposes additional variation upon experimental results. 

Angelini \textit{et al.} \cite{Angelini2011} studied the dynamics of a confluent epithelial cell sheet. They monitored 
cell motion over a broad range of length scales, time scales and cell densities. They found that 
with increasing cell density, the dynamics slow down. The log of the inverse self-diffusion coefficient
was found to have non-linear dependence on the cell density, which shares some vague analogy with the non-linear 
dependence of the log of the relaxation time on the inverse temperature. 
Even more interestingly, Angelini \textit{et al.} found
that not only the dynamics were slowing down but also they were increasingly more heterogeneous. They
estimated the dynamic correlation length and found that it increased with increasing cell density.

Garcia \textit{et al.} \cite{Garcia2015} studied a different confluent epithelial cell sheet. They also found a slowing
down of the dynamics upon increasing the cell density. They investigated dynamic heterogeneity and
determined the dynamic correlation length. Interestingly, upon increasing the cell density (which increased
during the duration of the experiment) the length exhibited a non-monotonic behavior, first increasing
and then decreasing with increasing density. Garcia \textit{et al.} also found a distinctly non-equilibrium
feature of active glassy dynamics; non-trivial equal time velocity correlations. We recall that in equilibrium
systems, either colloidal or molecular, equal-time velocity correlations are trivial; velocities of different
particles are uncorrelated. 
In contrast, Garcia \textit{et al.} 
determined that the length scale  characterizing these correlations also 
exhibits a non-monotonic dependence
on the cell density. Notably, the dynamic correlation length and the velocity correlation length were found to 
be correlated; their relation was found to be monotonic, in spite of the complex dependence of each length itself on time. 

More recently, Mongera \textit{et al.} \cite{Mongera2018} performed a more complex series of experiments. 
They focused on the important biological process of vertebrate body axis elongation. They identified and investigated 
an amorphous solidification process in which the cells become solid-like as they transition from mesodermal 
progenitor zone (MPZ) to presomitic mesoderm (PSM). The transition was monitored through 
the mean-square displacement, which was found to increase in a diffusive way in the the MPZ and
exhibit arrest in the PSM. They also studied mechanical response of both PSM and MPZ and identified
a yield stress, which is a commonly observed mechanical property of amorphous solids. Finally, they 
investigated active fluctuations and the role they play in the amorphous solidification. They found
that active fluctuations are strong in the MPZ and weak in the PSM, in analogy with thermal 
fluctuations in liquid and glassy phases. This finding led them to hypothesize that these active 
fluctuations play the role of a temperature. As we mentioned earlier, in our view the amorphous solidification
process uncovered by Mongera \textit{et al.} is a glass rather than a jamming transition, since it happens at
a non-vanishing level of the activity.

Colloidal systems have long served as a laboratory to observe and understand glassy 
dynamics~\cite{HunterWeeks,GokhaleAinP}. The reason is that many
colloidal experiments allowed workers to obtain significantly more information about the microscopic dynamics of
the colloidal systems than in atomic systems. The wealth of available information compensates for
the fact that in colloidal systems the slowing down is not as spectacular as in atomic systems and typically
only five or six decades of the change of the relaxation times can be observed. Colloidal systems consisting
of active Janus colloids (in which one part of the colloidal particle is covered with some kind of catalyst,
leading to a self-propelled motion) were one of the first synthetic active matter systems. Initially, the 
experiments focused on single-particle motion and then on moderately dense systems~\cite{Howse,Palacci,Buttinoni}. In the latter systems, 
clustering and phase separation of active colloidal particles with purely repulsive interactions can be 
observed~\cite{Buttinoni,Theurkauff}. More recently, some groups started investigating the structure and dynamics of dense active
colloidal systems~\cite{Klongvessa2019,Klongvessa2019b}. Although the details of the experiments are just emerging, it is clear that the dynamics of 
dense active colloidal systems exhibit classic signatures of glassy colloidal dynamics, with non-trivial dependencies of the microscopic 
dynamics upon changes in control parameters. We hope that further
work in this area will yield a wealth of information on the interplay between colloidal crowding and phoretic activity, since such experiments probe an simpler version of the more complex tissue dynamics. 

Finally, we mention a new experimental active matter system that belongs to the category of driven granular 
systems. For some time Dauchot's group has used macroscopic grains driven by shaking the plate on which they are 
placed as an experimental model system to study glassy dynamics and jamming phenomena~\cite{CoulaisSM}. Strictly speaking, this
driven system is active in the sense that it is an athermal system, devoid of any intrinsic dynamics and 
driven at the level of individual particles. However, since the drive is memory-less and the grains and the
motion are isotropic, the most appropriate theoretical and/or simulational model for this system is an
effective equilibrium system with thermal fluctuations. 
Recently, Dauchot's group introduced two new model systems of shaken grains. The first system consist of
monodisperse polar grains whose asymmetry leads to persistent motion under shaking~\cite{Deseigne}. These grains, therefore,
behave very much as self-propelled particles with ballistic short-time motion and effectively diffusive long-time motion. 
The system of polar grains is best modeled by models used for polar active matter~\cite{model-olivier,model-olivier2}.
More recently, a bidisperse version of the polar grain system was also introduced. In this system, crystallization is suppressed, and active glassy dynamics can be observed. 
Preliminary results again suggest important slowing down of the final diffusive motion accompanied
by an intermediate-time localization of individual particles. We hope that further analysis will investigate the presence of correlated motion or velocity correlations in this system.

\chapter{Oscillatory drive}

In this section we provide two examples of many-body interacting models driven out of equilibrium by harmonic time-dependent forces. The first model can be considered in the context of hysteretic phenomena in magnetic materials, whereas the second one is inspired by the physics of active epithelial tissues.

\section{A solvable model}

From what we discussed in the Sec.~\ref{sec:intro}, active matter should have all the characteristics 
related to macroscopic systems that underlie a statistical description, but none of the specifically equilibrium ones.
It is interesting then to study a problem of this kind that is completely solvable \cite{ivan}, and to see in detail what we may 
and what we may not say. 

Consider a mean-field XY model with Hamiltonian
\begin{equation}
H= \frac 1 N \sum_{i,j=1}^N \cos (\theta_i-\theta_j )
\label{hmf}
\end{equation}
We expect it to have a low-temperature phase that is magnetized along a direction:
\begin{equation}
\langle M_x+iM_y \rangle = \frac 1{2N} \sum_i \langle (\cos \theta_i + i \sin \theta_i) \rangle = M e^{i \Psi}
\label{magn}
\end{equation}
 all directions $\Psi$  being
equivalent, and a high temperature phase with zero magnetization $M=0$. 
Consider now this model in contact with a heat bath and perturbed  with an alternate field:
\begin{equation}
\dot \theta_i = -  \frac 1 N \sum_{j=1}^N  \sin (\theta_i-\theta_j ) -h\cos(\omega t)   \cos \theta_i + \eta_i 
\end{equation}
where $\eta_i$ are independent white noises with variance $2T$.
The dynamics is easy to solve for zero temperature, when all the angles are the same $\theta_i(t)=\theta(t)$:
\begin{equation}
\theta(t) =
2 \tan^{-1} \left(e^{-[\frac{h}{\omega} \sin(\omega t) + k]}\right)- 
  \frac{\pi}{2}
\label{cf}
\end{equation}
By considering all possible values of the integration constant $k$, we
conclude that at zero temperature solutions are possible in which the
total magnetization vector $M= e^{i \theta(t)}$ oscillates
around any possible angle.
 Unlike the case without field, these solutions are
no longer related by a continuous symmetry, and only the discrete
symmetries
$\theta \rightarrow -\theta$ and $\theta \rightarrow \pi-\theta$ remain.

\begin{figure}
\begin{center}
\includegraphics[width=10cm]{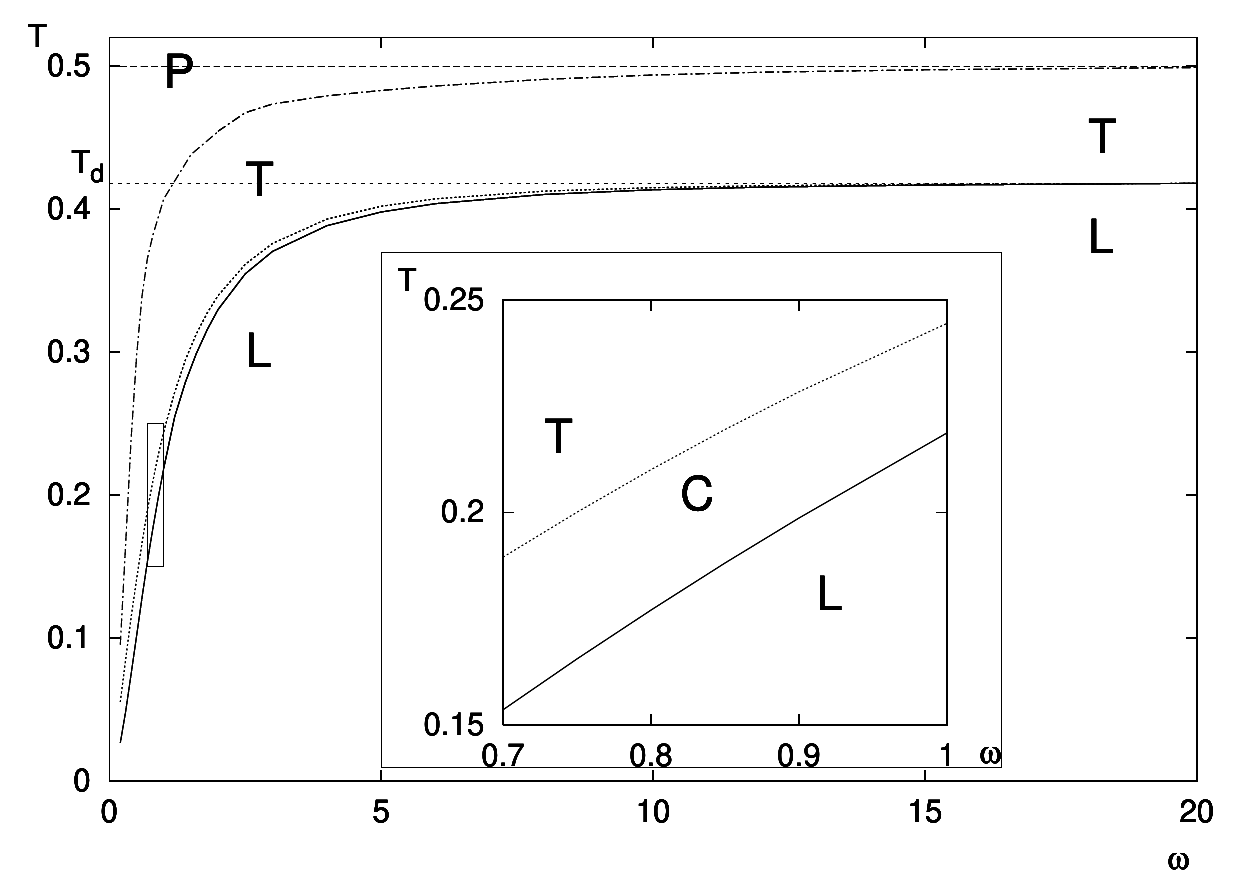}
\caption{
$T$-$\omega$ Phase diagram for the  model, $h=1$. }
\label{fig}
\end{center}
\end{figure}

When  we switch on temperature, the system may drift from an average angle to another.
But now, unlike the equilibrium case,  there is no symmetry that imposes that all average angles are equivalent, so that
we expect only some angles to be stable. Following Ref. \cite{ivan}, let us then solve the problem for all
temperatures. In terms of the mean field $M_x+iM_y  = \frac 1N \sum_j  (\cos \theta_j + i \sin \theta_j) $ we have,
for a single angle:
\begin{equation}
\dot \theta = -   M_x(t)  \sin \theta + M_y(t)  \cos \theta -h \cos(\omega t) \cos \theta + \eta
\label{ii}
\end{equation}
Using the Fokker-Plank equation associated with (\ref{ii}):
\begin{equation}\frac{dP}{dt} = 
\frac{\partial }{\partial \theta}\left[ T \frac{\partial }{\partial \theta}+
( M_x(t)  \sin \theta - M_y(t)  \cos \theta +h \cos(\omega t) )\right]P (\theta)  
\end{equation}
 we write an exact infinite system of equations for 
$c_n(t) \equiv \int d\theta \; P(\theta,t) \cos(n \theta) $, 
$s_n(t) \equiv  \int d\theta \; P(\theta,t) \sin(n \theta)$, $n=1,...$: 
\begin{eqnarray}
{\dot{s}}_n&=&-n^2Ts_n+\frac{1}{2}nc_1(s_{n-1}-s_{n+1})
+m (c_{n-1}+c_{n+1})\nonumber\\
{\dot{c}}_n&=&-n^2Tc_n+\frac{1}{2}nc_1(c_{n-1}-c_{n+1})
-m (s_{n-1}+s_{n+1})
\nonumber\\
{\dot{s}}_1&=&(\frac{1}{2}-T)s_1-\frac{1}{2}c_1s_2-\frac{1}{2}(s_1+h \cos(\omega t))c_2+\frac{1}{2}h \cos(\omega t)
\nonumber\\
{\dot{c}}_1&=&(\frac{1}{2}-T)c_1-\frac{1}{2}c_1c_2-\frac{1}{2}(s_1+h \cos(\omega t))s_2
\label{equa}
\end{eqnarray}
with $m(t)=n(s_1+h \cos(\omega t))/2$ and the mean-field closure $c_1(t)=M_x$ and $s_1(t)=M_y$.

\begin {figure}
\begin{center}
\input {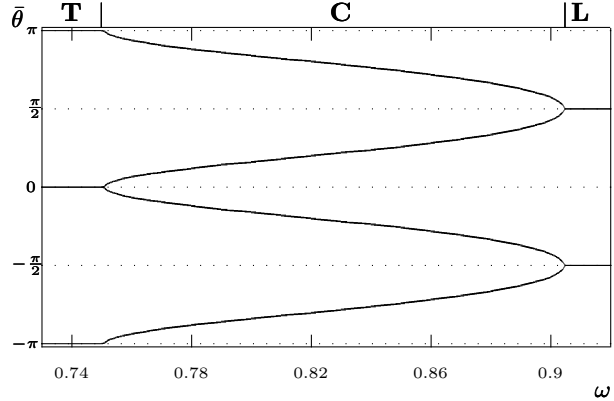}
\caption{
$\bar \theta$ as a function of $\omega$ 
for the  model, $h=1$, $T=0.2$. The transitions are second order.}
\label{fig2}
\end{center}
\end {figure}

The system  (\ref{equa}) may be easily solved (just a few modes are necessary)
for all values of $h,\omega, T$. 
The results, taken from \cite{ivan},  are presented in the figures \ref{fig} and \ref{fig2}.\\
 {\it (i) Paramagnetic}: The magnetization follows the field with a 
delay (hysteresis) and is zero on time-average.\\
 {\it (ii) Longitudinal $(\theta(t)=0 \; \; or \; \;\theta(t)=\pi)$}:
 the magnetization points  in the 
direction of the field: $M_\perp(t)=0$, but $\overline{M_h} \neq 0$.\\
 {\it  (iii) Transverse $(\bar \theta=\pi/2$ or $\bar \theta=-\pi/2)$:} 
the magnetization  has a non-zero component $M_\perp(t)\neq 0$ 
orthogonal to the field,
the component parallel to the field has zero time-average.\\
 {\it (iv) Canted $(0<\bar \theta <\pi/2$ or $\pi/2<\bar \theta <\pi)$:} 
The magnetization evolves around an oblique angle with the field's direction. The dependence of the angle 
on the frequency is shown in figure \ref{fig2}

Let us pause and reflect on what we have learned.
\begin{itemize}
\item The system allows for a description in terms of {\it macroscopic} variables. It has several
 transitions.
 Magnetizations have critical behavior around the transitions, with mean-field critical exponents.
In these senses, it is just like any thermodynamic system. \\

\item Is there a simple general principle that allows us to guess which is the average angle that is selected for 
a given amplitude and frequency? Or, in other words, do the chosen angles optimize the power injected, the entropy
production or something we may interpret as a free-energy? The answer seems negative. The only
thing that it seems we may say is that the chosen average angle maximizes the probability of being there: a  truism.

\end{itemize}

\section{A model for active matter with oscillatory drive}

As mentioned above, oscillatory driving forces represent a generic way to drive a physical towards a non-equilibrium state. An oscillatory drive is qualitatively distinct from the type of active forces (mainly, self-propulsion) discussed in Sec.~\ref{active} above, and this section is dedicated to an example of an active model devised to schematically describe the dynamics of epithelial tissue~\cite{elsen}, where the active forces driving the dynamics are indeed sinusoidal in time.

In this model, a confluent tissue is modeled as soft repulsive particles at large density, but now the source of activity is given by spontaneous (`active') fluctuations of the particle volume. Experimental observations in dense epithelial tissues indeed suggest that individual cells undergo relatively large volume fluctuations (up to 20\%) that appear almost periodic in time, with a very low frequency~\cite{volume}. The idea of the model is thus to remove all kinds of other active forces and deal with volume fluctuations alone, in order to understand the physics associated to such an oscillatory drive in a dense amorphous material. The main outcome of the model is to display a non-equilibrium phase transition between a solid-arrested phase at low amplitude of the oscillatory driving force, to a fluid-flowing phase at larger amplitude. Interestingly, the solid-fluid phase transition appears discontinuous and is not associated to a glassy slowdown of the type discussed in Secs.~\ref{twobaths} and \ref{glassyself} above for models of  self-propelled particles. 

We consider a dense suspension of $N$ soft circular particles at 
zero temperature in a two-dimensional square box of linear size $L$ with
periodic boundary conditions.
The interaction between the particles is modelled by a short-ranged 
repulsive harmonic potential, similar to jammed foams~\cite{durian}:
$V(r_{ij})=\frac{\epsilon}{2} \left( 1- r_{ij}/ \sigma_{ij} 
\right)^2 H(\sigma_{ij}-r_{ij})$,
where $r_{ij}=|\vec r_i - \vec r_j|$,  $\sigma_{ij}=(\sigma_i + \sigma_j)/2$, 
with $\sigma_i$ and $\vec r_i$ the diameter and position of particle $i$, 
respectively.
The energy scale of the repulsive force is set by $\epsilon$,
and $H(x)$ is the heaviside function, defined such that $H(x \ge 0)=1$.
In the overdamped limit, the dynamics of each particle is described 
by a Langevin equation:
\begin{equation}
\xi\frac{d\vec r_i}{dt}=-\sum_{j\neq i}\frac{\partial V(r_{ij})}{\partial 
\vec r_j},
\label{eq:motion}
\end{equation} 
where $\xi$ is a friction coefficient.
The dissipation timescale is $\tau_0= \xi \sigma_0^2/\epsilon$, 
where $\sigma_0$ sets the particle diameter. Physically, $\tau_0$ is the typical timescale for a system described by Eq.~(\ref{eq:motion}) to come at rest without forcing.

\begin{figure}
\begin{centering}
\includegraphics[width=8.5cm]{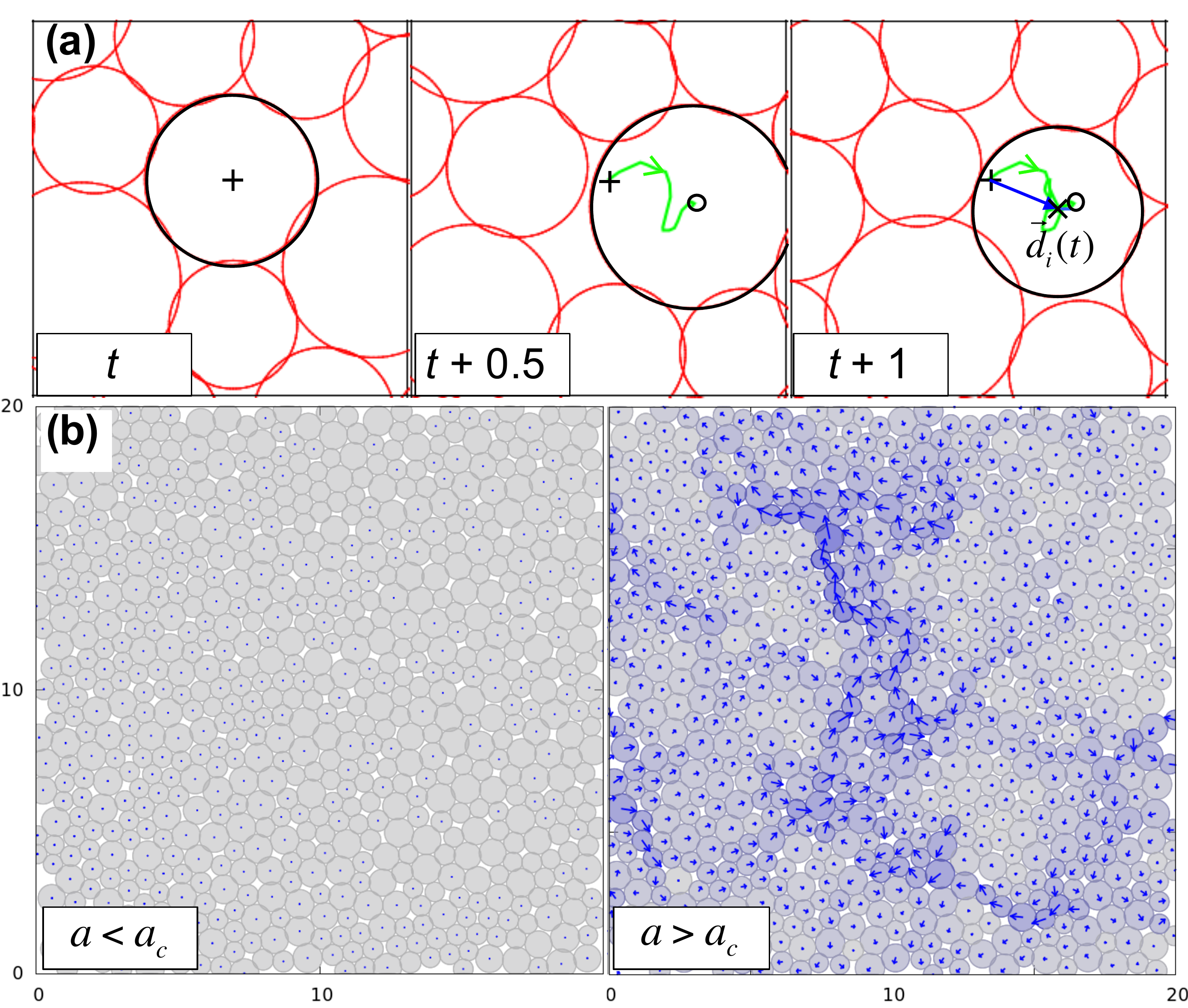}
\par\end{centering}
\protect\caption{(a) Snapshots of the system over one cycle of 
active deformation. 
The green curve is the trajectory of the highlighted particle 
during one cycle. The blue arrow represents its 
displacement after one cycle $\vec d_i(t)= \vec r_i(t+1)-\vec r_i(t)$.
(b) One-cycle displacement map $\vec d_i(t)$ in steady state for the 
disordered solid phase ($a=0.047<a_c \approx 0.049$, left) and in the 
fluid phase ($a=0.051>a_c$, right).
In the solid phase, particles approximately return 
to their position after each cycle. In the fluid, 
there are regions of large displacements where irreversible rearrangements
take place.  The transition between reversible and irreversible 
phases at $a_c$ is discontinuous.
\label{fig:model}}
\end{figure}

The system is driven out of equilibrium by oscillating the diameter of each particle around its mean value $\sigma_i^0$, 
as shown in Fig.~\ref{fig:model}(a):
\begin{equation}
\sigma_i(t) = \sigma_i^0 \left[1 + a\cos\left( \omega t +
\psi_i\right)\right],
\label{eq:oscillation}
\end{equation} 
where $T = 2 \pi /\omega$ is the period of oscillation which is used as time unit, and $a$ is an adimensional parameter which quantifies the 
intensity of the activity.
We impose very slow oscillations, $T \gg \tau_0$, 
such that the system is always located near an energy minimum
and inertial and hydrodynamic effects can be neglected. 
Specifically, we use $T = 820 \tau_0$. 
The average diameters $\sigma_i^0$ are drawn from a 
bidisperse distribution of diameters $0.71\sigma_0$ and $\sigma_0$ with 3:2 
proportion, in order to prevent crystallization. 
We use $\sigma_0$ as unit length.
We have introduced in Eq.~(\ref{eq:oscillation}) a random phase $\psi_i$ for each particle 
to constrain the total area fraction $\phi=\sum_{i} \frac{\pi\sigma_i^2(t)}{4L^2}$ to be strictly constant in time. 
The case with $\psi_i \equiv 0$ would correspond to affine compressions 
and expansions, 
which would then amount to studying the 
rheological response of the jammed solid forced at large scale, 
not an active material forced locally. 
We consider jammed systems with $\phi=0.94$, as appropriate for 
confluent tissues. Most simulations were 
performed with a very large system of $N=16000$ particles
(typically $L \approx 100\sigma_0$). 
For each $a$ value, we prepare fully random systems and apply 
the periodic perturbation until the system has reached steady state, 
either arrested or flowing. We then perform 
steady state measurements using averaging over time and initial conditions (in the flowing
phase), or over initial conditions (in the arrested phase). 

Figure~\ref{fig:model}(a) highlights the trajectory of a particle
during one period. The one-cycle displacement,
$\vec d_i(t) = \vec r_i(t+1) - \vec r_i(t)$, is
 shown in Fig.~\ref{fig:model}(a). Collecting
the displacement of all particles we obtain the steady state one-cycle 
displacement map shown in Fig.~\ref{fig:model}(b) for both arrested 
and flowing phases. 
In the arrested phase, displacements are all very 
small and particles approximately return to the same 
position after each cycle, without undergoing configurational change.
On the other hand, at large activity  we observe regions of very large 
displacements where irreversible particle rearrangements
occur within one cycle. These local plastic events are spatially 
disordered, and they coexist with regions where displacements
are smaller: the dynamics is spatially heterogeneous.
Clearly, Fig.~\ref{fig:model} indicates the
existence of an arrested phase where particles do not move for small
$a$, and of a flowing phase for large $a$ 
where irreversible rearrangements take place during each cycle.


\begin{figure}
\begin{center}
\includegraphics[height=6.cm]{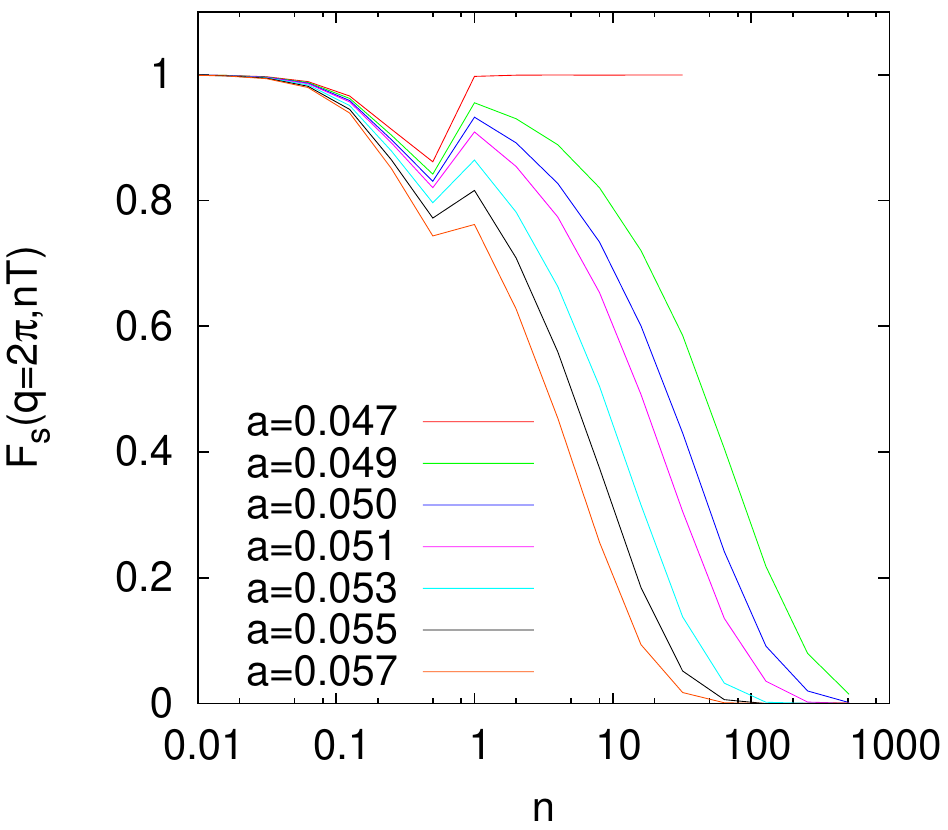}
\includegraphics[height=6.cm]{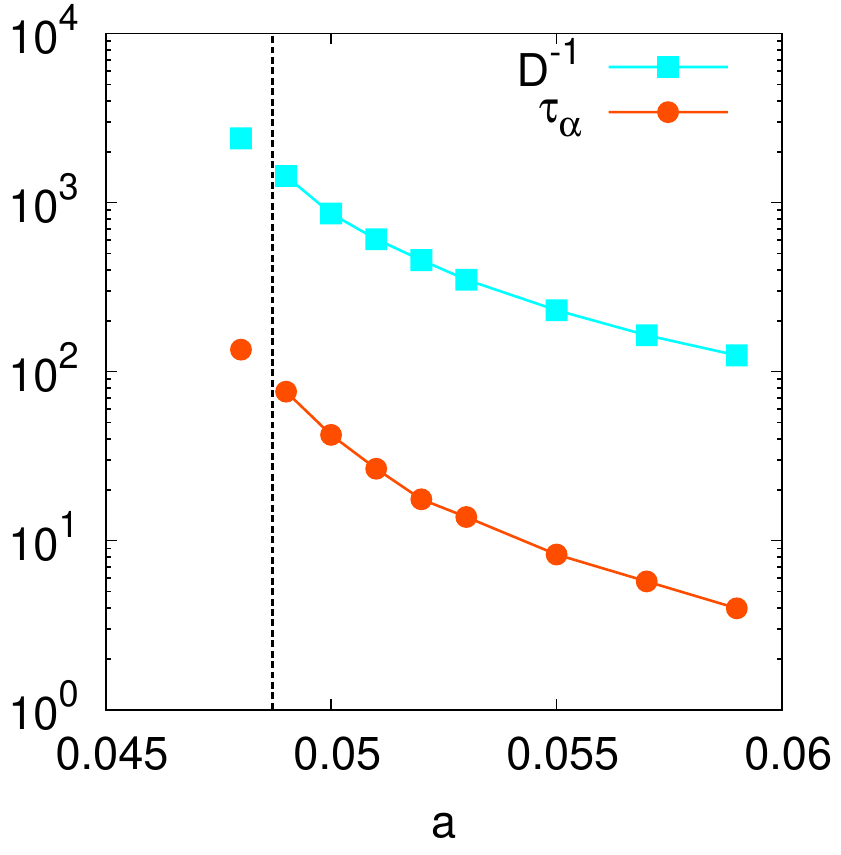}
\end{center}
\caption{Left: Time dependence of the self-intermediate scattering function as the amplitude of the forcing is decreased. Right: Discontinuous divergence of the relaxation time $\tau_\alpha$ and vanishing of the diffusion constant $D$ at the non-equilibrium first-order transition at $a_c$.} 
\label{fig:timescale}
\end{figure}

Turning to the long-time dynamics, we can measure the the mean-squared displacements of the particles to extract the self-diffusion constant $D$ of the particles. It is found that a finite diffusion constant exists for $a > a_c$, but the diffusivity vanishes at small forcing amplitude $a<a_c$. 
We also measure the self-intermediate scattering function $F_s(q,n)$, see Eq.~(\ref{eq:self}), now measured as a function of the number of cycles, $n$. The results are shown in Fig.~\ref{fig:timescale}. This correlation function rapidly decays to zero at large amplitude, defining a finite structural relaxation $\tau_\alpha$ as $F_s(q,n=\tau_\alpha)=1/e$ in the flowing phase, which becomes infinite in the arrested phase.  

In Fig.~\ref{fig:timescale} we also report $D^{-1}$ and $\tau_\alpha$ as a function of activity $a$.
Both measures of long-time dynamics increase modestly by about $1$ decade as  $a \to a_c^+$, and they do not diverge.
In addition, just below $a_c$, we find that the flowing phase can be `metastable' for a long time of order $30 \tau$ 
before suddenly evolving towards the arrested phase. 
Within this metastability window, long-time dynamical properties can be measured 
and we plot $D^{-1}$ and $\tau$ for this metastable liquid phase as 
isolated points in Fig.~\ref{fig:timescale}, which appear 
as the continuation of data at $a>a_c$. 
These observations confirm that both timescales do not diverge 
at $a_c$ and illustrate the first-order nature of the phase transition at $a_c$. 

The proper analogy with the physics of glassy materials is not with the glass transition itself, but rather with the yielding transition discussed in Sec.~\ref{sec:rheology} which is also found to be discontinuous~\cite{takeshi} under periodic driving conditions. Physically, volume fluctuations can be seen as a slow driving force, and fluidization occurs when that force exceeds a threshold, as for yielding~\cite{yieldRMP}, the only difference being that the force acts on a local (for active systems) rather than a global (for rheology) scale. The transition exhibited by the self-deforming particle model thus qualifies as an `active yielding transition'~\cite{elsen}. 

In the above numerical model, the volume oscillations were taken as purely periodic with a very low but fixed frequency. In the opposite limit where frequencies are widely distributed or changing with time, the driving becomes random and resembles the random fluctuations provided by thermal noise, thus leading to ordinary equilibrium-like glassy dynamics. This case was explored in the original paper~\cite{elsen}, and was recently revisited in more detail numerically~\cite{atsushirecent}. In this recent work, fluctuations in the volume of the particles was introduced to model in a rather crude manner conformational changes of proteins filling the cytoplasm of bacteria. The simulations confirm that volume fluctuations with arbitrary (rather than monochromatic) time-dependent fluctuations yields thermal-like glassy dynamics and a non-equilibrium glass transition phenomenon, as opposed to the sharp discontinuous transition obtained for an oscillatory drive.

\end{document}